\RequirePackage{fix-cm}
\documentclass[smallextended]{svjour3}       % onecolumn (second format)
\smartqed  % flush right qed marks, e.g. at end of proof
\usepackage{graphicx}
\usepackage{amsmath}
\usepackage{amsfonts}
\usepackage{amssymb}
\usepackage{mathtools}
\usepackage{physics}
\usepackage{siunitx}

\usepackage{adjustbox}
\usepackage{epstopdf}
\usepackage{float}
\usepackage{graphicx}
\usepackage{rotating}
\usepackage{subcaption} %for subfiguresm, cannot be just together with subfig
\usepackage{tikz}
\usepackage{pgfplots}
\pgfplotsset{compat=1.18}

\usepackage{booktabs}
\usepackage{longtable}
\usepackage{threeparttable}
\usepackage{multirow}

\usepackage{xcolor}
\usepackage[ruled]{algorithm2e}
\usepackage[labelfont={bf}]{caption}
\usepackage{enumerate}
\usepackage{lastpage}
\usepackage{listings}
\usepackage{ulem}
\usepackage{hyperref}
\usepackage{authblk}
\usepackage{bbm}

\usepackage{verbatim}
\usepackage{url}
\usepackage{etoolbox}

\usepackage{draftwatermark}
\SetWatermarkText{}  %add text
\newcommand{\blue}[1]{\textcolor{black}{#1}}

\providetoggle{showpt}
\settoggle{showpt}{false}

\DeclarePairedDelimiter\floor{\lfloor}{\rfloor}

\newcommand{\E}{\mathbb{E}}
\newcommand{\Prob}{\mathbb{P}}

\DeclareMathOperator{\cvar}{CVaR}

\DeclareMathOperator{\ess}{ess}

\newcommand{\R}{\mathbb{R}}

\newcommand{\naturalnumbers}{\mathbb{N}}

\date{\today} % The date
\allowdisplaybreaks[1]

\newtheorem{axiom}{Axiom}

\newenvironment{conditions*}
{\par\vspace{\abovedisplayskip}\noindent
	\tabularx{\columnwidth}{>{$}l<{$} @{${}={}$} >{\raggedright\arraybackslash}X}}
{\endtabularx\par\vspace{\belowdisplayskip}}

\journalname{arXiv}
\begin{document}

\title{Quantifying the degree of risk aversion of spectral risk measures%\thanks{Grants or other notes
%about the article that should go on the front page should be
%placed here. General acknowledgments should be placed at the end of the article.}
}
%\subtitle{Do you have a subtitle?\\ If so, write it here}

%\titlerunning{Short form of title}        % if too long for running head

\author{E. Ruben van Beesten}

%\authorrunning{Short form of author list} % if too long for running head

\institute{E. Ruben van Beesten \at
              Econometric Institute, Erasmus University Rotterdam, the Netherlands\\
              Department of Industrial Economics and Technology Management, Norwegian University of Science and Technology, Trondheim, Norway. \\
              \email{vanbeesten@ese.eur.nl}            %  \\
%             \emph{Present address:} of F. Author  %  if needed
}

\date{Received: date / Accepted: date}
% The correct dates will be entered by the editor

\maketitle

\begin{abstract}
    \color{black} %\color{blue}
    This paper introduces a quantitative notion of the degree of risk aversion of spectral risk measures. We define a family of degree functionals $D_p$, $p \in \R$, characterized by three axioms governing normalization, mixtures, and continuity. The resulting degrees admit representations in terms of both the Kusuoka measure and the dual utility function, leading to connections with generalized means, the Gini coefficient, and an Arrow–Pratt-type measure of curvature. We further relate the parameter $p$ to the tail behavior of losses through generalized Pareto distributions, which provides an interpretation of the choice of $p$ and a basis for calibrating spectral risk measures according to their desired treatment of different tail behaviors. The degree functional is consistent with several dominance relations between dual utility functions, and the full degree profile uniquely identifies a spectral risk measure. Finally, we extend the degree functional to law-invariant coherent risk measures through their minimal Kusuoka representations.
    \color{black}
    %This paper introduces the first axiomatic quantification of the \textit{degree of risk aversion} for spectral and law-invariant coherent risk measures. This notion formalizes the idea that some risk measures are ``more risk-averse'' than others. To this end, \blue{we} propose a \textit{degree functional} on the space of spectral risk measures that is uniquely defined by two axioms: a normalization on the space of CVaRs and a linearity axiom. \blue{We} present two formulas for the degree functional and discuss several properties and interpretations. Finally, \blue{we} extend the definition to law invariant coherent risk measures.
% \PACS{PACS code1 \and PACS code2 \and more}
 % 90C15: stochastic programming, 91B05: risk models
\end{abstract}

%\newpage

%% =======================
%% MAIN BODY BELOW
%% =======================

\section{Introduction} \label{sec:introduction}

\color{black} %\color{blue}
Spectral risk measures are widely used to incorporate risk aversion into stochastic optimization models (see, e.g., \cite{duan2025day,ge2024sorel,jin2025managing,kim2024spectral,lu2025spectral,moghimi2025beyond,wu2024multistage} for recent applications and \cite{guigues2012sddp} for an example by the late Werner R\"{o}misch). While different spectral risk measures reflect different attitudes toward risk, there is currently no principled way to quantify how risk-averse a given spectral risk measure is, or to compare different spectral risk measures in terms of their degree of risk aversion. 
Such a quantification would provide a common scale for comparing and interpreting risk measures, and could facilitate their calibration according to a desired level of conservatism.

This paper introduces a quantitative notion of the \textit{degree of risk aversion} of spectral risk measures. To this end, we define a family of degree functionals $D_p$, parametrized by $p \in \R$, that assign to each spectral risk measure a value between zero and one, with higher values corresponding to higher degrees of risk aversion. The functionals are characterized by three axioms: a normalization that assigns the $\alpha$-conditional value-at-risk ($\mathrm{CVaR}_\alpha$) the degree $\alpha$, a mixing axiom that determines how degrees are aggregated when spectral risk measures are mixed, and a continuity axiom that extends this principle to continuous mixtures. The parameter $p$ determines how strongly the degree reflects the more or less risk-averse components of such mixtures.

The degree functional is inspired by the literature on quantifying risk aversion in theories of choice under uncertainty. In the expected-utility framework of von Neumann and Morgenstern \cite{neumann1947theory}, the most prominent quantitative measures are the local measures of absolute and relative risk aversion introduced by Arrow \cite{arrow1965aspects} and Pratt \cite{pratt1964risk},
%In this framework, various quantifications of the degree of risk aversion have been proposed based on the (dis)utility function. The most prominent examples are the measures of absolute and relative risk aversion introduced by Arrow and Pratt \cite{arrow1965aspects,pratt1964risk}. These measures are local, in the sense that they quantify risk aversion at a given wealth or cost level, 
while several authors have proposed global measures that summarize risk aversion over the entire domain \cite{caballe2007stochastic,chateauneuf2005more,eisenhauer2006integral}. 
Related notions have also been developed in the dual theory of choice under risk by Yaari \cite{yaari1987dual} for the resulting spectral risk measures. In particular, local Arrow-Pratt-type indices for spectral risk measures have been studied in \cite{eeckhoudt2022dual,wachter2013consistent},  while \cite{brandtner2015decision} investigate comparative risk aversion between spectral risk measures. 
%Other authors have developed risk and probability premia and higher-order measures of dual risk attitudes \cite{eeckhoudt2020risk}. 
%These contributions provide local measures of the intensity of risk aversion or qualitative orderings for comparing decision makers.

Our objective is different: we seek a \textit{global}, numerical degree that can be assigned to each spectral risk measure individually. Specifically, we construct a family of degree functionals $D_p$ that assign each spectral risk measure a value between zero and one indicating its degree of risk aversion. Due to the aforementioned normalization, the degree has a direct interpretation on a common CVaR-scale, allowing spectral risk measures to be compared numerically even when existing comparative-risk-aversion relations do not order them. In this sense, the proposed degree functional complements existing local and comparative notions of risk aversion in the dual framework and can be viewed as a dual counterpart to global measures of risk aversion developed for expected utility.

%Spectral risk measures, on the other hand, are rooted in Yaari’s dual theory of rational choice under uncertainty [39]. While risk aversion is well understood qualitatively in this dual framework, no analogous quantification of the degree of risk aversion is known. The degree functional introduced in this paper can therefore be viewed as a dual counterpart to global measures of risk aversion developed for expected utility.

We derive two complementary representations of the degree functional $D_p$. The first expresses the degree in terms of a $p$-generalized mean of the CVaR levels underlying the spectral risk measure through its Kusuoka representation. This representation gives a direct interpretation of the parameter $p$: smaller values of $p$ make the degree more sensitive to the more risk-averse components of the mixture, whereas larger values place relatively more emphasis on the less risk-averse components. The second representation expresses $D_p$ in terms of the dual utility function associated with the spectral risk measure and leads to further interpretations. In particular, for $p=1$ the degree coincides with the Gini coefficient of the dual utility function, while for $p=-1$ it can be related explicitly to an Arrow–Pratt-type measure of its curvature.

The parameter $p$ also admits an interpretation in terms of the tail behavior of the losses under consideration. We show that, for $p>-1$, the degree $D_p(\rho)$ is uniquely determined by the value that $\rho$ assigns to a generalized Pareto loss with extreme-value index $\xi=-p$. Since generalized Pareto distributions describe exceedances over high thresholds in extreme-value theory, this result suggests choosing $p=-\xi$ when losses have extreme-value index $\xi<1$. It also provides a concrete calibration interpretation: rather than specifying a spectral risk measure directly through its risk spectrum or dual utility function, one may specify how conservatively losses with different tail behaviors should be evaluated and use the corresponding degrees to calibrate the risk measure.

We further establish several structural properties of the degree functional. In particular, we study its consistency with dominance relations between the dual utility functions associated with spectral risk measures. We show that dominance in relative-convexity order implies a higher degree of risk aversion for every $p$, while stochastic dominance implies the same ordering for appropriate ranges of $p$. These results show that the quantitative ordering induced by $D_p$ is consistent with natural qualitative notions of one spectral risk measure being more risk-averse than another. Finally, we show that the collection of degrees $D_p(\rho)$, $p\in\mathbb{N}$, uniquely determines the spectral risk measure $\rho$.

Finally, we extend the degree functional from spectral risk measures to the more general class of law-invariant coherent risk measures. Using the minimal Kusuoka representation, we define the degree of a law-invariant coherent risk measure as the supremum of the degrees of the spectral risk measures appearing in its representation. This extension agrees with the original degree functional on the spectral subclass and preserves the normalization on CVaR.

The remainder of the paper is structured as follows. Section~\ref{sec:preliminaries} introduces the probabilistic setting and reviews law-invariant coherent and spectral risk measures. Section~\ref{sec:construction} provides the axiomatic construction of the degree functional and derives its two representations. Section~\ref{sec:interpretations} discusses several interpretations of the degree functional.
Section~\ref{sec:calibration} relates the parameter $p$ to the tail behavior of losses and discusses the resulting calibration of spectral risk measures. Section~\ref{sec:properties} establishes several of its structural properties, while Section~\ref{sec:dominance} studies its consistency with dominance relations between dual utility functions. 
Section~\ref{sec:coherent} extends the degree functional to law-invariant coherent risk measures, and Section~\ref{sec:conclusion} concludes. Some of the longer proofs are presented in the appendix.

\color{black}

\color{black} %\color{blue}

\section{Preliminaries} \label{sec:preliminaries}

This section introduces the probabilistic setting and notation used
throughout the paper and reviews the two classes of risk measures considered:
law-invariant coherent risk measures and spectral risk measures.

\subsection{Setting and notation}

We consider the space $\mathcal{Z} = L^\infty(\Omega, \mathcal{F}, \Prob)$ of bounded random variables on an atomless probability space $(\Omega, \mathcal{F}, \Prob)$, where random variables $Z \in \mathcal{Z}$ represent costs or losses, so smaller realizations are more favorable. The reason for restricting to $L^\infty(\Omega, \mathcal{F}, \Prob)$ rather than $L^q(\Omega, \mathcal{F}, \Prob)$ for some finite $q \in [1,\infty)$ is that among all $q \in [1,\infty]$, the choice $q=\infty$ leads to the largest collection of (finite-valued) risk measures $\rho : \mathcal{Z} \to \R$. In particular, it uniquely allows for including the essential supremum, defined as
\begin{align*}
    \ess \sup (Z) := \sup\{ z \in \R \ : \ \Prob\{Z \geq z \} > 0\}, \quad Z \in \mathcal{Z},
\end{align*}
in our analysis (as $\ess \sup (Z) < \infty$ for all $Z \in \mathcal{Z} = L^\infty(\Omega, \mathcal{F}, \Prob)$), which will play the role of the ``most risk-averse'' spectral risk measure in our framework. The main results generalize straightforwardly to a setting with $q < \infty$; also see Remark~\ref{remark:finite_q}. Throughout, $\E$ denotes expectation with respect to $\Prob$, i.e., $\E[Z] := \int_\Omega Z\, d\Prob$, $Z \in \mathcal{Z}$.

\subsection{Law-invariant coherent and spectral risk measures}

On the space $\mathcal{Z}$, we consider two collections of \textit{risk measures}, i.e., functionals on $\mathcal{Z}$: law-invariant coherent risk measures and spectral risk measures. Although spectral risk measures are our main interest in this paper, we first define the more general notion of law-invariant coherent risk measures.

We define $\bar{\mathcal{R}}$ as the collection of all risk measures $\rho : \mathcal{Z} \to \R$ that are \textit{law invariant} and \textit{coherent}. Law invariance means that if $Z_1,Z_2 \in \mathcal{Z}$ have the same law (i.e., the same cdf), then $\rho(Z_1) = \rho(Z_2)$. Moreover, $\rho$ is coherent if it satisfies the four axioms defined in \cite{artzner1999}: monotonicity, positive homogeneity, translation equivariance, and subadditivity. Every law-invariant coherent risk measure $\rho \in \bar{\mathcal{R}}$ admits a Kusuoka representation \cite{kusuoka2001law,shapiro2013kusuoka} of the form
\begin{align}
    \rho(Z) = \sup_{\mu \in \mathcal{M}_\rho} \int_{[0,1]} \cvar_\alpha(Z) d\mu(\alpha), \quad Z \in \mathcal{Z}, \label{eq:kusuoka_coherent}
\end{align}
where $\mathcal{M}_\rho$ is a subset of the collection $\mathcal{P}([0,1])$ of probability measures on $([0,1],\mathcal{B}_{[0,1]})$, and $\cvar_\alpha$ denotes the risk measure \textit{conditional value at risk} \cite{acerbi2002,rockafellar2002}, which is defined for $\alpha \in [0,1)$ as
\begin{align}
    \cvar_\alpha(Z) = \int_\alpha^1 F_Z^{-1}(t) dt, \quad Z \in \mathcal{Z}, \label{eq:def_cvar}
\end{align}
with $F_Z^{-1}$ the quantile function (i.e., the generalized inverse of the cumulative distribution function (cdf)) of $Z$, and for $\alpha = 1$ as $\cvar_1(Z) = \lim_{\alpha \uparrow 1} \cvar_\alpha(Z) = \ess \sup Z$.\footnote{Note that for $\alpha = 0$, we have $\cvar_0 = \E$.}  The Kusuoka representation shows that every law-invariant coherent risk measure can be interpreted as a worst-case mixture of CVaRs. Although the set $\mathcal{M}_\rho$ satisfying \eqref{eq:kusuoka_coherent} is not unique, \cite{pichler2012uniqueness} shows that there exists a unique minimal set $\mathcal{M}_\rho$. 

Next, we define $\mathcal{R}$ as the collection of all \textit{spectral} risk measures \cite{acerbi2002spectral} on $\mathcal{Z}$. This consists of all law-invariant coherent risk measures $\rho \in \bar{\mathcal{R}}$ that are also \textit{comonotone additive}, meaning that if $Z_1,Z_2 \in \mathcal{Z}$ are comonotonic, then $\rho(Z_1 + Z_2) = \rho(Z_1) + \rho(Z_2)$. We emphasize that $\mathcal{R} \subseteq \bar{\mathcal{R}}$, i.e., every spectral risk measure is law-invariant coherent. For spectral risk measures $\rho \in \mathcal{R}$, the set $\mathcal{M}_\rho$ in the Kusuoka representation \eqref{eq:kusuoka_coherent} reduces to a singleton $\{\mu_\rho\}$, so we can write
\begin{align}
    \rho(Z) = \int_{[0,1]} \cvar_\alpha(Z) d\mu_\rho(\alpha), \quad Z \in \mathcal{Z}. \label{eq:kusuoka_spectral}
\end{align}
Moreover, spectral risk measures admit a spectral representation of the form
\begin{align}
    \rho(Z) = \int_0^1 F_Z^{-1}(u) dw_\rho(u), \qquad Z \in \mathcal{Z}, \label{eq:spectral}
\end{align}
where $w_\rho$ is a cdf on $[0,1]$ that is convex and satisfies $w_\rho(0) = 0$. The function $w_\rho$ is called the \textit{dual utility function} associated with $\rho$ and it represents how much each quantile of the distribution of $Z$ is weighted by $\rho \in \mathcal{R}$ in \eqref{eq:spectral}. Its name originates from Yaari's dual theory of choice under risk \cite{yaari1987dual}, where spectral risk measures arise as a numerical representation of preference relations satisfying a set of axioms that are, in a specific sense, dual to the axioms in the (primal) expected utility theory of von Neumann and Morgenstern \cite{neumann1947theory}. In this dual framework, risk-aversion of preferences corresponds to convexity of the dual utility function $w_\rho$. We denote the left derivative $\varphi_\rho := w_\rho^{(l)}$ as the \textit{risk spectrum} associated with $\rho$.\footnote{\blue{Our definition of the risk spectrum here is slightly more general than its typical definition as the derivative of $w_\rho$ under the assumption that $w_\rho$ is differentiable.}} The two representations \eqref{eq:kusuoka_spectral} and \eqref{eq:spectral} are related through the following standard identities which will be used throughout the paper.

\begin{lemma}\label{lemma:identities}
    Let $\rho \in \mathcal{R}$ be a given spectral risk measure and consider its Kusuoka representation \eqref{eq:kusuoka_spectral} and its spectral representation \eqref{eq:spectral}. Then, the dual utility function $w_\rho$ can be expressed as\footnote{\blue{Here, at $t=1$, the fraction $\frac{1-\alpha}{1-\alpha}$ is taken to equal 1 at $\alpha = 1$.}}
    \begin{align}
        w_\rho(t) = \int_{[0,t]} \frac{t-\alpha}{1-\alpha} d\mu_\rho(\alpha), \quad t \in [0,1]. \label{eq:w_as_rho}
    \end{align}
    Moreover, the right and left derivatives of $w_\rho$, written as $w_\rho^{(r)}$ and $w_\rho^{(l)} = \varphi_\rho$ are given by
    \begin{align}
        w_\rho^{(r)}(t) = \int_{[0,t]}(1-\alpha)^{-1} d\mu_\rho(\alpha), \quad t \in [0,1),\label{eq:w_r_as_mu}
    \end{align}
    and
    \begin{align}
         w_\rho^{(l)}(t) = \int_{[0,t)}(1-\alpha)^{-1} d\mu_\rho(\alpha) +\mu_\rho(\{1\}) \chi_{(t = 1)}, \quad t \in (0,1], \label{eq:phi_as_mu}
    \end{align}
    where $\chi_{(t=1)} = +\infty$ if $t = 1$ and $\chi_{(t=1)} = 0$ otherwise, and we use the convention $0 \cdot (+\infty) = 0$.    
\end{lemma}
\begin{proof}
    Equation \eqref{eq:w_as_rho} is standard and follows, e.g., by integrating  both sides of Equation~(5) in \cite{pichler2012spectralriskmeasuresadaptions}. This can be written equivalently as
    \begin{align*}
        w_\rho(t) = \int_{[0,1]} \frac{(t-\alpha)_+}{1-\alpha} d\mu_\rho(\alpha), \quad t \in [0,1].
    \end{align*}
    To find the left and right derivatives, we separate the end point $t = 1$ explicitly:
    \begin{align*}
        w_\rho(t) = \int_{[0,1)} \frac{t-\alpha}{1-\alpha} d\mu_\rho(\alpha) + \mathbbm{1}_{(t=1)} \mu_\rho(\{1\}), \quad t \in [0,1].
    \end{align*}
    The integrand $t \mapsto \frac{t-\alpha}{1-\alpha}$ in the first term has left and right derivatives $\frac{\mathbbm{1}_{\alpha < t}}{1-\alpha}$ and $\frac{\mathbbm{1}_{\alpha \leq t}}{1-\alpha}$, respectively. The second term has left derivative $\mu_\rho(\{1\})\chi_{(t=1)}$. Hence, taking the integral of the first term's right and left derivatives yields $w^{(r)}(t) = \int_{[0,1]} \frac{\mathbbm{1}_{\alpha \leq t}}{1-\alpha} d\mu_\rho(\alpha) = \int_{[0,t]} (1-\alpha)^{-1} d\mu_\rho(\alpha)$, $t \in [0,1)$, and $w^{(l)}(t) = \linebreak \int_{[0,1]} \frac{\mathbbm{1}_{\alpha < t}}{1-\alpha} d\mu_\rho(\alpha) + \mu_\rho(\{1\})\chi_{(t=1)} = \int_{[0,t)} (1-\alpha)^{-1} d\mu_\rho(\alpha) + \mu_\rho(\{1\})\chi_{(t=1)}$, $t \in (0,1]$.
    \qed
\end{proof}

\color{black}

\section{Axiomatic construction of $D_p$} \label{sec:construction}

\color{black} %\color{blue}
This section presents the axiomatic construction of the degree functional $D_p$. We introduce three axioms in Section~\ref{subsec:axioms}, and find corresponding representations in Section~\ref{subsec:formulas}.
\color{black}

\subsection{Axioms} \label{subsec:axioms}

\color{black} %\color{blue}
As a starting point, we take the collection of conditional value-at-risk measures $\cvar_\alpha \in \mathcal{R}$, $\alpha \in [0,1]$. This collection provides a natural benchmark because its parameter $\alpha$ already has a clear interpretation as a level of risk aversion: increasing $\alpha$ shifts attention toward increasingly adverse outcomes. Moreover, its endpoints range from $\cvar_0 = \E$, representing risk neutrality, to $\cvar_1 = \ess \sup$, representing arguably the most risk-averse spectral risk measure. We therefore adopt the normalization $D_p(\cvar_\alpha) = \alpha$ as our first axiom, ensuring we preserve the natural ordering and scale of the CVaR family and making the degree $D_p$ directly interpretable as a CVaR-equivalent confidence level.
\color{black}

\begin{axiom}[Normalization] \label{ax:normalization}
    The degree of $\cvar_\alpha$ equals its parameter $\alpha$, i.e.,
    \begin{align}
        D_p(\cvar_\alpha) = \alpha, \qquad \alpha \in [0,1]. \label{eq:ax_normalization}
    \end{align}
\end{axiom}

\color{black} %\color{blue}
To extend $D_p$ to non-CVaR spectral risk measures, we recall that by the Kusuoka representation \eqref{eq:kusuoka_spectral}, every spectral risk measure $\rho$ can be interpreted as a mixture of CVaRs. Axiom~\ref{ax:normalization} determines the degree of every individual CVaR, but does not determine how these degrees should be combined when CVaRs are mixed. There is no uniquely compelling aggregation rule: one may wish to give relatively more influence to either the more or the less risk-averse components. We therefore consider a parametrized family of mixing principles. We take $\alpha=1$, corresponding to $\cvar_1=\ess \sup$, as the maximally risk-averse reference point. Accordingly, $1-D_p(\rho)$ represents the distance of $\rho$ from this reference point. For a fixed $p \in \R$, Axiom~\ref{ax:linearity} requires linearity after transforming this distance by $x\mapsto x^p$, with the logarithmic transformation used for $p=0$. That is, we require linearity of the function $f_p : \R \to \bar{\R}$, defined by
\begin{align}
    f_p(x) := \begin{cases}
        (1 - x)^p, &\text{if } p \neq 0, \\
        \log(1 - x), &\text{if } p = 0,
    \end{cases}
    \label{eq:f_p}
\end{align}
with the conventions that $(0)^p = +\infty$ if $p < 0$  and $\log(0) = -\infty$. The parameter $p \in \R$ thereby controls the relative influence of the more and less risk-averse components of the mixture.

\color{black} %\color{blue}
\begin{axiom}[$p$-linear mixing] \label{ax:linearity}
    The transformation $s_p := f_p \circ D_p$ of the degree functional $D_p$, with $f_p$ defined in \eqref{eq:f_p}, is linear in convex combinations. That is, if $\rho_1,\ldots,\rho_n \in \mathcal{R}$ and $\lambda_1,\ldots,\lambda_n >0$ such that $\sum_{i=1}^n \lambda_i = 1$, then
    \begin{align}
        s_p\left(\sum_{i=1^n} \lambda_i \rho_i\right) = \sum_{i=1}^n \lambda_i s_p(\rho_i). \label{eq:ax_linearity}
    \end{align}
\end{axiom}
\color{black}

%The mixing axiom above is indeed flexible: the weighting of each risk measure out of those being mixed depends on the parameter $p$. Specifically, if $p=1$, then $D_p$ is linear itself, so every risk measure has equal weight. Lower values of $p$ imply more emphasis on the risk measure $\rho_i$ with a \textit{higher} value of $D_p(\rho_i)$, while higher values of $p$ imply more emphasis on the risk measures $\rho_i$ with \textit{lower} values of $D_p(\rho_i)$. 

%One benefit of choosing a parametrized mixing axiom of this particular $p$-linear form is that the resulting degree functional $D_p$ takes the form of a $p$-generalized mean (as we will show in Section~\ref{subsec:formulas}). This facilitates its interpretation and allows for the derivation of interesting properties, particularly for the cases $p=1$, $p=0$ and $p=-1$, which correspond to the arithmetic, geometric and harmonic mean (see Sections~\ref{subsec:formulas} and \ref{sec:properties}). The question what parameter choice for $p$ is most meaningful in different situations will be explored in Section~\ref{sec:properties}.

\begin{comment}
\begin{remark}
    Linearity of $s_p$ is equivalent to linearity of $h_p \circ D_p$, where $$h_p(\alpha) = \begin{cases}
        -p^{-1}\big( (1 - \alpha)^p - 1 \big), &\text{if } p \neq 0, \\
        \lim_{p \to 0} -p^{-1}\big( (1 - \alpha)^p - 1 \big), &\text{if } p = 0.
    \end{cases}$$ 
    The function $h_p$ is occasionally useful, plotting it for different values of $p$ can help with intuitions. \label{ftn:h_p}
\end{remark}
\end{comment}

\color{black} %\color{blue}
While Axiom~\ref{ax:linearity} deals with finite mixtures of CVaRs, we need one more axiom to deal with continuous mixtures. To this end, we define a continuity axiom. The axiom implies that for $\rho$ with continuous $\mu_\rho$, its associated $D_p(\rho)$ is found as the limit of $D_p(\rho_{\mu_n})$, where $\mu_n$ are finitely discrete measures in $\mathcal{P}([0,1])$ that converge weakly to $\mu_\rho$. Since for $p \leq 0$ the function $f_p$ from Axiom~\ref{ax:linearity} is unbounded as $\alpha \uparrow 1$, we avoid pathological sequences that would contradict Axioms~\ref{ax:normalization} and \ref{ax:linearity} by restricting ourselves to sequences for which $\int f_p d\mu_n \to \int f_p d\mu$.  

\begin{axiom}[Continuity] \label{ax:continuity}
    Let $\{\mu_n\}_{n=1}^\infty$ be a sequence of probability measures in $\mathcal{P}([0,1])$ and let $\mu \in \mathcal{P}([0,1])$ be given. Suppose that
    \begin{enumerate}[(a)]
        \item $\mu_n \Rightarrow \mu$, i.e., $\mu_n$ weakly converges to $\mu$, and
        \item $\int f_p d\mu_n \to \int f_p d\mu$.
    \end{enumerate}
    Then, $D_p(\rho_{\mu_n}) \to D_p(\rho_\mu)$. 
\end{axiom}

No more axioms are needed. All spectral risk measures are CVaR mixtures; Axiom~\ref{ax:normalization} fixes $D_p$ on CVaR, Axiom~\ref{ax:linearity} extends this to finite CVaR mixtures, and Axiom~\ref{ax:continuity} to continuous CVaR mixtures.

\color{black}

\subsection{\blue{Representations of $D_p$}} \label{subsec:formulas}

\blue{We now show that there is a uniquely defined degree functional} $D_p$ that satisfies Axioms~\ref{ax:normalization}--\ref{ax:continuity}. \blue{We derive two equivalent representations in Theorems~\ref{thm:r_formula_mu} and \ref{thm:r_formula_w}. We will postpone interpretation of the representations to Section~\ref{sec:interpretations}. The first representation is in terms of the measure $\mu_\rho$ from the Kusuoka representation \eqref{eq:kusuoka_spectral} of $\rho \in \mathcal{R}$.} 

\begin{theorem} \label{thm:r_formula_mu}
    For every $p \in \R$, there is a unique function $D_p : \mathcal{R} \to \R$ that satisfies Axioms~\ref{ax:normalization}--\ref{ax:continuity}. Moreover, it can be represented as
    \begin{align}
        D_p(\rho) = 1 - \blue{M_p(1 - A_{\mu_\rho})}, \label{eq:r_formula_mu}
    \end{align}
    where $\mu_\rho$ is the measure from the Kusuoka representation  \eqref{eq:kusuoka_spectral} of $\rho$, and $M_p(1-A_{\mu_\rho})$ denotes the $p$-generalized mean of $1-A_{\mu_\rho}$, where $A_{\mu_\rho} \sim \mu_\rho$, i.e.,
    \begin{align*}
         \blue{M_p(1 - A_{\mu_\rho})} = \begin{cases}
             \left( \int_{[0,1]} (1 - \alpha)^p d\mu_\rho(\alpha) \right)^{1/p}, &\text{if } p \neq 0, \\
             \exp{ \int_{[0,1]} \log(1 - \alpha) d\mu_\rho(\alpha) }, &\text{if } p = 0.
         \end{cases} 
    \end{align*}
\end{theorem}
\begin{proof}
    See Appendix~\ref{sec:proofs} 
\end{proof}

\color{black} %\color{blue}
\begin{remark}\label{remark:finite_q}
Although we work throughout with spectral risk measures defined on $\mathcal{Z} := L^\infty(\Omega,\mathcal{F},\Prob)$, the results remain valid for spectral risk measures defined on $L^q(\Omega, \mathcal{F},\Prob)$, $1\leq q<\infty$, provided that the risk measure is finite-valued (and hence continuous \cite{kaina2009convex}) on $L^q$. Indeed, the restriction of such a risk measure to $L^\infty$ is a spectral risk measure with the same dual utility function $w_\rho$ and the same Kusuoka measure $\mu_\rho$. Theorem~\ref{thm:r_formula_mu} may therefore be applied to this restriction, and its conclusion depends only on $\mu_\rho$, rather than on the choice of the space on which the risk measure is defined. Conversely, since $L^\infty$ is dense in $L^q$, every $Z\in L^q$ can be approximated in $L^q$ by its bounded truncations $Z^{(m)}:=(-m)\vee(Z\wedge m)$. The $L^q$-continuity of $\rho$ then implies $\rho(Z^{(m)})\to \rho(Z)$, so any argument involving variables in $L^q$ follows from the corresponding bounded-variable argument by truncation. Thus, the choice $q=\infty$ is made only for notational and technical convenience and does not affect the results.
\end{remark}
\color{black}

For future reference, we give a formal definition of the degree functional.

\begin{definition}[Degree functional] \label{def:D_p}
    For every $p \in \R$, the functional \linebreak $D_p \ : \mathcal{R} \to \R$ defined in \eqref{eq:r_formula_mu} is called the ($p$th) \textit{degree (of risk aversion) functional}. Moreover, for every $\rho \in \mathcal{R}$, we refer to $D_p(\rho)$ as its \textit{$p$th degree (of risk aversion)}. 
\end{definition}

Another representation of $D_p$ can be found in terms of the dual utility function $w_\rho$ corresponding to $\rho$ or its left derivative $\varphi_\rho$.

\begin{theorem} \label{thm:r_formula_w}
    The degree functional $D_p$ can be represented as
    \begin{align}
        D_p(\rho) = \begin{cases}
            1 - \left[ (p+1) \int_0^1 (1 - t)^p dw_\rho(t) \right]^{1/p} &p > -1 \text{ and } p \neq 0, \\
            1 - \exp{\int_0^1 \log(1-t) dw_\rho(t) + 1}, &p = 0, \\
            1 - \left[ \varphi_\rho(1) \right]^{-1}, &p = -1, \\
            %\blue{1 - \left[ \varphi_\rho(1) - (p+1) \int_0^1 (1-t)^p (\varphi_\rho(1) - \varphi_\rho(t)) dt \right]^{1/p},} &\blue{p < -1.}
            \blue{1 - \left[ \int_{[0,1]} (1-t)^{p+1} d\varphi_\rho(t) \right]^{1/p},} &\blue{p < -1.}
        \end{cases}
        \label{eq:r_formula_w}
    \end{align}
    for all $\rho \in \mathcal{R}$.
\end{theorem}
\color{black} %\color{blue}

\begin{proof}
    See Appendix~\ref{sec:proofs}.
\end{proof}

\color{black} %\color{blue}

\section{Interpretations of $D_p$} \label{sec:interpretations}

In this section, we provide interpretations of the degree functional $D_p$ for different values of $p$. We first provide general interpretations of $D_p$ based on its two characterizations in Theorems~\ref{thm:r_formula_mu} and \ref{thm:r_formula_w}. Then, we provide special representations of $D_p$ for the special cases $p=1$ and $p=-1$, which allow for intuitive interpretations. 
%Finally, we derive a close connection between $D_p$ and generalized Pareto distributions from extreme-value theory, which can guide the choice of the order parameter $p$.

\subsection{General interpretations of $D_p$}

Theorem~\ref{thm:r_formula_mu} expresses $D_p(\rho)$ in terms of the Kusuoka measure $\mu_\rho$ associated with $\rho$, which describes $\rho$ as a mixture of CVaRs. If $A_{\mu_\rho} \sim \mu_\rho$, then $D_p(\rho) = 1 - M_p(1 - A_{\mu_\rho})$. Thus, $1 - D_p(\rho)$ is the $p$-generalized mean of the distances $1-\alpha$ between the CVaR levels occurring in the mixture and the maximally risk-averse level $\alpha = 1$, which serves as a reference point in the representation. Equivalently, $D_p(\rho)$ aggregates the degrees $\alpha$ of the constituent CVaRs, with the precise manner of aggregation determined by $p$. If $\mu_\rho$ places more mass on high CVaR levels, then the generalized mean of $1 - \alpha$ becomes smaller and $D_p(\rho)$ becomes larger. This is consistent with the interpretation of a mixture involving more upper-tail-sensitive CVaRs as being more risk-averse. 

The parameter $p$ determines how the different CVaR levels in the mixture are weighted. For $p=1$, the degree is simply the arithmetic mean of the CVaR levels:
\begin{align}
    D_1(\rho) = 1 - M_1(1 - A_{\mu_\rho}) = M_1(A_{\mu_\rho}) = \int_{[0,1]} \alpha \, d\mu_\rho(\alpha). \label{eq:D_1_integral}
\end{align}
For smaller values of $p$, the generalized mean assigns relatively more importance to small values of $1-\alpha$, and hence to CVaR components with $\alpha$ close to one. The resulting degree therefore becomes increasingly sensitive to the most risk-averse components of the mixture. Special cases are $p=0$, for which $M_0$ is the geometric mean, and $p=-1$, for which $M_{-1}$ is the harmonic mean. Conversely, larger values of $p$ place relatively more importance on large values of $1-\alpha$, corresponding to the less risk-averse CVaR components. In this sense, $p$ controls whether the degree primarily reflects the most or the least conservative components of the risk measure.

Theorem~\ref{thm:r_formula_w} gives a complementary interpretation in terms of the dual utility function $w_\rho$. For $p > -1$, the formula can be written as a transformation of a moment of $1 - T_\rho$, where $T_\rho$ is a random variable with cdf $w_\rho$. Since $w_\rho$ is convex, it lies below the identity function and the distribution induced by $w_\rho$ places relatively more mass toward the right endpoint of $[0,1]$. The degree $D_p$ quantifies the extent of this concentration near one: shifting probability mass toward larger values of $T_\rho$ reduces the relevant generalized moment of $1-T_\rho$ and consequently increases $D_p(\rho)$. The parameter $p$ determines which part of this concentration is emphasized. In particular, decreasing $p$ makes the degree progressively more sensitive to mass situated very close to the endpoint $1$. 

A structural break occurs at $p = -1$, where $D_p(\rho)$ is determined fully by the left derivative $\varphi_\rho(1)$ of $w_\rho$ at one.\footnote{This value often appears in bounds related to spectral risk measures; see, e.g., \cite{vanbeesten2020convex,vanbeesten2024convex}.} For $p < -1$, the integral in the representation is no longer with respect to $w_\rho$, but with respect to its left derivative $\varphi_\rho$. Consequently, as $p$ decreases below $-1$, $D_p$ increasingly reflects how sharply the dual utility function $w_\rho$ bends upward near one, i.e., how rapidly the risk spectrum $\varphi_\rho$ increases toward the most adverse outcomes.
%The interpretation of $D_p(\rho)$ becomes more extreme: where for $p > -1$ it is a measure of the location of the probability mass induced by $w_\rho$, with more mass near $\alpha=1$ leading to higher values, for $p < -1$ it is a measure of the location of the \textit{curvature} of $w_\rho$, with more curvature around $\alpha = 1$ leading to higher values.

%\blue{For $p > -1$, Theorem~\ref{thm:r_formula_w} represents $D_p$ as a function of the expectation of $(1-T)^p$ (or $\log(1-T)$ if $p=0$), where $T$ is a random variable with cdf $w_\rho$. The more probability mass is concentrated around $1$, the larger the resulting value of $D_p(\rho)$. Thus, $D_p(\rho)$ is a measure of how heavily $w_\rho$ concentrates probability mass around $1$. The lower the value of $p$, the higher the relative weighting of mass very close to one. The limit occurs as $p = -1$, where $D_p(\rho)$ is determined fully by the left derivative $\varphi_\rho(1)$ of $w_\rho$ at one. Note that as $w_\rho$ is convex, $\varphi_\rho(1)$ is also the maximum value the risk spectrum attains. This value often appears in bounds related to spectral risk measures; see, e.g., \cite{vanbeesten2020convex,vanbeesten2024convex}. Finally, if $p < -1$, the integral is with respect to the risk spectrum itself. Thus, $D_p$ measures how strongly the risk spectrum becomes more concentrated on increasingly extreme losses as one moves into the upper tail. }

\subsection{Gini and Wasserstein interpretations for $p=1$}

For the special case $p=1$, Theorem~\ref{thm:r_formula_w} simplifies to
\begin{align}
    D_1(\rho) = 2 \int_0^1 t \, dw_\rho(t) - 1. \label{eq:D_1_rho}
\end{align}
Interpreting $w_\rho$ as a Lorenz curve, we can interpret $D_1(\rho)$ as the \textit{Gini coefficient} \cite{gini1912variabilita} associated with $w_\rho$. This concept is often used in economics to measure income inequality \cite{cowell2011measuring}; it quantifies the distance from a certain income distribution to a hypothetical distribution with complete income equality. In our context, it measures the deviation between the cdf $w_\rho$, which assigns probability mass disproportionately to the right end of the interval $[0,1]$, and the cdf $t \mapsto t$ of a uniform distribution on $[0,1]$. 

\color{black} %\color{blue}
\begin{definition}[Gini coefficient]    
    Let $v : [0,1] \to [0,1]$ be an admissible Lorenz curve, i.e., a convex function satisfying $v(0) = 0$ and $v(1) = 1$. Then, the \textit{Gini coefficient} associated with  $v$ is given by
    \begin{align*}
        G(v) := 2 \int_0^1 (x - v(x)) dx.
    \end{align*}
\end{definition}

\begin{proposition}[Gini interpretation for $p=1$] \label{prop:gini}
    Let $p=1$ and let $\rho \in \mathcal{R}$ be given. Then, $D_1(\rho)$ is the Gini coefficient associated with $w_\rho$, i.e.,
    \begin{align*}
        D_1(\rho) = G(w_\rho) =  2\int_0^1 (t - w_\rho(t))dt. % = 1 - 2 \int_0^1 w_\rho(t) dt 
    \end{align*}
\end{proposition}
\color{black}
\begin{proof}
    Using integration by parts for Stieltjes integrals, we have
    \begin{align*}
        \int_0^1 t \, dw_\rho(t) &= 1 \cdot w_\rho(1) - 0 \cdot w_\rho(0) - \int_0^1 w_\rho(t) dt = 1 - \int_0^1 w_\rho(t) dt.
    \end{align*}
    Substituting this into \eqref{eq:D_1_rho} yields
    \begin{align*}
        D_1(\rho) &= 2 \int_0^1 t \, dw_\rho(t) - 1 =  2(1 - \int_0^1 w_\rho(t) dt) - 1 = 1 - 2\int_0^1 w_\rho(t) dt \\
        &= 2\int_0^1 (t - w_\rho(t))dt.
    \end{align*}
    This concludes the proof. \qed
\end{proof}

\begin{figure}
    \centering
    \usetikzlibrary{intersections}
\usetikzlibrary{patterns}
\usepgfplotslibrary{fillbetween}

\begin{tikzpicture}
  \begin{axis}[
      axis lines=middle,
      xlabel={$t$},
      ylabel={},
      xmin=0, xmax=1.05,
      ymin=0, ymax=1.05,
      xtick={0,1},
      ytick={0,1},
      domain=0:1,
      samples=100,
      axis line style={->},
      thick,
      width=8.8cm,
      height=6.8cm,
      every axis x label/.style={at={(ticklabel* cs:1.05)}, anchor=west},
      every axis y label/.style={at={(ticklabel* cs:1.05)}, anchor=south},
      clip=false,
    ]

    % Risk-neutral line: w(t) = t
    \addplot [name path=T, black, dashed, domain=0:1] {x}
      node [pos=0.8, above left] {$u(t)=t$};

    % Convex curve: more risk-averse spectral risk measure
    \addplot [name path=R, thick, blue, domain=0:1] {x^4}
      node [pos=0.5, below right] {$w_\rho(t)$};

    % Shaded area between w(t)=t and w_rho(t)=t^3
    \addplot [
      blue!30,
      pattern=north east lines,
      pattern color=blue!40,
    ] fill between[of=T and R];

    % Annotate Gini coefficient
    \node at (axis cs:0.55,0.3) [align=center]
      {$\frac{1}{2}D_1(\rho)$};

  \end{axis}
\end{tikzpicture}
    \caption{Illustration of the interpretation of $D_1(\rho)$ as a Gini coefficient for a spectral risk measure $\rho$ with dual utility function $w_\rho(t) = t^4$.}
    \label{fig:gini}
\end{figure}

\begin{corollary} \label{cor:wasserstein}
    Let $\rho \in \mathcal{R}$ be a spectral risk measure. Then,
    \begin{align*}
        D_1(\rho) = 2 \, W_1(\Prob_{w_\rho}, \Prob_u),
    \end{align*}
    where $\Prob_{w_\rho}$ is the probability measure on $[0,1]$ induced by the cdf $w_\rho$, $\Prob_u$ is the probability measure on $[0,1]$ induced by the uniform cdf $u(t) = t$, and $W_1$ is the type-1 Wasserstein distance.% (also know as the ``Kantorovich distance'' or the ``earth mover's distance''). 
\end{corollary}
\begin{proof}
    Write $u(t) = t$ for the cdf of the uniform distribution on $[0,1]$. Since $w_\rho(0) = 0$, $w_\rho(1) = 1$ and $w_\rho$ is convex, it follows that $w_\rho(t) \leq t = u(t)$ for all $t \in [0,1]$. That is, $w_\rho$ first-order stochastically dominates $u$. By Proposition~3.2 in \cite{deAngelis2021whyWasserstein}, this implies that $2 W_1(\Prob_{w_\rho}, \Prob_u) = 2(\int_0^1 t \, dw_\rho(t) - \int_0^1 t \, du(t)) = 2(\int_0^1 t \, dw_\rho(t) - 1/2) = 2 \int_0^1 t \, dw_\rho(t) - 1 = \blue{D_1}(\rho)$, where the last equality follows from \eqref{eq:D_1_rho}. \qed
\end{proof}

\color{black} %\color{blue}
Proposition~\ref{prop:gini} and Corollary~\ref{cor:wasserstein} are illustrated in Figure~\ref{fig:gini}. The value $D_1(\rho)$ is twice the area between the graph of $w_\rho$ and of $t \mapsto t$. This area achieves its minimum value of zero if $w_\rho(t) = u(t) = t$, which corresponds to $\rho = \cvar_0 = \E$, and its maximum value of one if $w_\rho(t) = 0$, $t \in [0,1)$, and $w_\rho(1) = 1$, which corresponds to $\rho = \cvar_1 = \text{ess sup}$. Note that these values correspond to the extreme values from Proposition~\ref{prop:bounds}.

The Gini interpretation reveals that $D_1$ is essentially the dual analogue to the index of global risk aversion in \cite{eisenhauer2006integral} for expected utility functionals. Moreover, it shows that $D_p$ is a measure of convexity of $w_\rho$. Indeed, in Yaari's dual theory of choice under risk \cite{yaari1987dual}, risk-aversion is equivalent to convexity of $w_\rho$. So, besides measuring the degree of risk aversion of $\rho$, the degree functional $D_1$ equivalently measures the ``degree of convexity'' of $w_\rho$, or of its curvature. This connects $D_1$ to classical quantifications of risk aversion for (primal) expected (dis)utility $\E[u(Z)]$, such as the Arrow-Pratt measure of absolute risk aversion \cite{pratt1964risk}. The latter is defined, for disutility functions $u$, as $A_u(z) = u''(z)/u'(z)$, and can be interpreted as a local measure of the curvature of the disutility function $u$ at loss level $z$. While the Arrow-Pratt measure quantifies \textit{local} curvature, the degree functional $D_1$ measures \textit{global} curvature of $w_\rho$ over its entire interval.

\color{black}

\subsection{Arrow-Pratt interpretation for $p=-1$}

\color{black} %\color{blue}
For $p=-1$, the degree functional $D_{-1}$ is closely related to the natural extension (see \cite{eeckhoudt2022dual,wachter2013consistent}) of the Arrow-Pratt measure of absolute risk aversion \cite{arrow1965aspects,pratt1964risk} to dual utility functions, defined as
\begin{align}
    A_{w_\rho}(t) = \frac{w_\rho''(t)}{w_\rho'(t)}, \quad t \in [0,1].\label{eq:def_arrow_pratt}
\end{align} 
This functional measures the \textit{local} curvature of $w_\rho$ at $t$. The following result shows that $D_{-1}(\rho)$ can be expressed in terms of the integral of $A_{w_\rho}$ over its domain (which is the dual analogue of the \textit{cumulative absolute risk aversion function} from \cite{nielsen2005monotone} for expected utility functionals). This shows that $D_{-1}$ is a \textit{global} measure of the curvature of $w_\rho$. 

\begin{proposition} \label{prop:arrow-pratt}
    Let $\rho \in \mathcal{R}$ be given and assume that $w_\rho$ is twice differentiable and that $\varphi_\rho(0) > 0$. Then,
    \begin{align}
        D_{-1}(\rho) = 1 - \frac{e^{-\int_0^1 A_{w_\rho}(t) dt}}{\varphi_\rho(0)}, \label{eq:D_-1_arrow}
    \end{align}
    where $A_{w_\rho}$ is the functional from \eqref{eq:def_arrow_pratt}.
\end{proposition}
\begin{proof}
    Note that $A_{w_\rho}(t) = \frac{d}{dt} \log(w_\rho'(t)) = \frac{d}{dt} \log(\varphi_\rho(t))$. Thus,
    \begin{align*}
        \int_0^1 A_{w_\rho}(t) dt = \int_0^1 \frac{d}{dt} \log(\varphi_\rho(t)) dt = \log(\varphi_\rho(1)) - \log(\varphi_\rho(0)) = \log\left(\frac{\varphi_\rho(1)}{\varphi_\rho(0)}\right).
    \end{align*}
    It follows that $\varphi_\rho(1) = \varphi_\rho(0) e^{\int_0^1 A_{w_\rho}(t) dt}$. The result now follows from substituting this expression into $D_{-1}(\rho) = 1 - 1/\varphi_\rho(1)$ from Theorem~\ref{thm:r_formula_w}.
    \qed
\end{proof}

%Proposition~\ref{prop:arrow-pratt} shows that, under differentiability conditions on $w_\rho$, the degree functional $D_{-1}(\rho)$ can be expressed in terms of the integral over the extended Arrow-Pratt measure $A_{w_\rho}$ over the interval $[0,1]$, i.e., in terms of the cumulative proportional curvature of the dual utility function $w_\rho$, together with the initial marginal weight $\varphi_\rho(0)$. 

\section{Tail behavior and calibration} \label{sec:calibration}

%\blue{Finally, we show that when $p > -1$, $D_p(\rho)$ is determined by the value that (a truncated version of) $\rho$ assigns to random variables that have a generalized Pareto distribution with shape parameter $-p$.}

The parameter $p$ determines how the degree functional aggregates the different levels of risk aversion represented in a spectral risk measure. In this section, we provide a complementary interpretation of $p$ in terms of the tail behavior of the losses being evaluated. In particular, we establish a connection between $D_p(\rho)$ and the value that $\rho$ assigns to generalized Pareto losses, whose shape parameter characterizes tail behavior in extreme-value theory. This connection provides guidance for the choice of $p$ and, more broadly, a way to calibrate spectral risk measures based on their desired treatment of losses with
different tail behaviors.

\color{black} %\color{blue}
\begin{theorem} \label{thm:equivalence}
   %Let $\rho_1, \rho_2 \in \mathcal{R}$ with $D_p(\rho_1) = D_p(\rho_2)$ be given. 
   Let $p > -1$ be given. For $\theta > 0$, let $Z_{p,\theta}$ be a random variable with a generalized Pareto distribution with shape parameter $-p$, scale parameter $1/\theta$, and location parameter $0$, i.e., $Z_{p,\theta}$ has cdf
   \begin{align*}
       F_{p,\theta}(z) = \begin{cases}
           1 - (1 - \theta p z)^{1/p},\quad \text{ for } 0 \leq z \leq (p\theta)^{-1}, &\text{if } p > 0, \\
           1 - e^{-\theta z},\ \quad \quad\qquad \text{ for } z \geq 0, &\text{if } p = 0, \\
           1 - (1 - \theta p z)^{1/p},\quad \text{ for } z \geq 0, &\text{if } -1 < p < 0.
       \end{cases} 
   \end{align*}
   Then, for every $\rho \in \mathcal{R}$, define\footnote{The need for this extension of $\rho$ is that $\rho \in \mathcal{R}$ was defined on the space $\mathcal{Z}$ of \textit{bounded} random variables, while for $p \leq 0$, $Z_{p,\theta}$ is unbounded.}
   \begin{align}
       \bar{\rho}(Z_{p,\theta}) := \lim_{m\to \infty} \rho(Z_{p,\theta} \wedge m), \label{eq:rho_truncated}
   \end{align}
   Then, $D_p(\rho)$ is uniquely determined by $\bar{\rho}(Z_{p,\theta})$. In particular,
   \begin{align}
       D_p(\rho) = \begin{cases}
           1 - \left[ (p+1) (1 - p \theta \bar{\rho}(Z_{p,\theta}) \right)]^{1/p}, &\text{if } p \neq 0, \\
           1 - \exp{1 - \theta \bar{\rho}(Z_{0,\theta})}, &\text{if } p = 0.
       \end{cases} 
       \label{eq:D_formula_Z}
   \end{align}
   %\begin{align*}
   %    \rho_1(Z_{p,\theta}) = \rho_2(Z_{p,\theta}).
   %\end{align*}
\end{theorem}
\begin{proof}
    See Appendix~\ref{sec:proofs}.
\end{proof}

Theorem~\ref{thm:equivalence} shows that $D_p(\rho)$ is uniquely determined by the value that the extension $\bar{\rho}$ of $\rho$ attains for a generalized Pareto distribution with shape parameter $-p$. This connects the degree functional $D_p$ to extreme value theory. The Pickands–Balkema–de Haan theorem \cite{balkema1974residual,pickands1975statistical} states that for random variables $Y$ satisfying mild conditions, for a very large threshold $u \in \R$, the distribution of the exceedance $Y - u$ conditional on $Y \geq u$ is approximately generalized Pareto distributed. The shape parameter $\xi = -p$ is the extreme-value index that characterizes the asymptotic tail behavior of the distribution of $Y$, with positive values of $\xi$ indicating heavy, polynomial-type tails and negative values indicating the absence of a tail (i.e., a finite upper point).  Theorem~\ref{thm:equivalence} therefore suggests that $D_{-\xi}$ is the degree functional that captures the relevant aspect of $\rho$ when the loss faced has extreme-value index $\xi$. In particular, for all risk measures $\rho \in \mathcal{R}$ with the same value of $D_{-\xi}(\rho)$ their extensions $\bar{\rho}$ agree on the value of $\bar{\rho}(Z_{{-\xi},\theta})$ for all $\theta > 0$. All in all, this suggests that $p=-\xi$ is the appropriate choice of the parameter $p$ when facing losses with extreme-value index $\xi < 1$.\footnote{\blue{Note that $\xi < 1$ is precisely the condition that guarantees that $Y$ has a finite mean.}}

\begin{remark}[Risk measure calibration]
     Theorem~\ref{thm:equivalence} can be used to calibrate a spectral risk measure $\rho$ on the family of generalized Pareto distributions. Rather than specifying the risk spectrum or dual utility function directly, one may specify how conservatively losses with different tail behaviors should be evaluated. For a generalized Pareto loss with extreme-value index $\xi<1$, such an assessment determines the corresponding degree $D_{-\xi}(\rho)$. Assessments at several values of $\xi$ therefore provide a risk-aversion profile that can be used to calibrate the parameters of a chosen family of spectral risk measures. For example, $p=1$ corresponds to a generalized Pareto distribution with $\xi=-1$, which is  uniform, while $p=0$ corresponds to the limiting case $\xi=0$, which is exponential; specifying $D_1(\rho)$ and $D_0(\rho)$ can therefore be interpreted as specifying the desired degree of conservatism toward bounded uniform losses and exponentially tailed losses, respectively. Note that these assessments cannot be chosen independently: by Proposition~\ref{prop:D_p_limits} below, $D_p$ is non-increasing in $p$, so if $\xi_1 < \xi_2 < 1$, then $D_{-\xi_1}(\rho) \leq D_{-\xi_2}(\rho)$. Thus, the framework both provides an interpretable way to calibrate a spectral risk measure from its desired treatment of losses with different tail behaviors and imposes consistency restrictions on such assessments.
\end{remark}

\color{black}

\section{Properties of $D_p$} \label{sec:properties}

\blue{In this section, we derive properties of the degree functional $D_p$. First, we determine its range, prove monotonicity in $p$, and derive its limiting values. Second, we show that every spectral risk measure is fully determined by its degree profile.}

\subsection{\blue{Range, monotonicity, and limiting values}}

\blue{First, we show that the range of $D_p$ is the interval $[0,1]$.}
\begin{proposition} \label{prop:bounds}
    For every $p \in \R$, it holds that
    \begin{align}
        D_p(\rho) \in [0,1], \qquad \rho \in \mathcal{R}, \label{eq:D_range}
    \end{align}
    and the bounds are attained by  $D_p(\E) = 0$ and $D_p(\ess \sup) = 1$.
\end{proposition}
\begin{proof}
    By Theorem~\ref{thm:r_formula_mu}, it suffices to show that $\blue{M_p(1 - A_{\mu_\rho})} \in [0,1]$ to prove the first claim. This follows from the fact that the generalized mean of a random variable is bounded by the bounds of the support of that random variable, which is the interval $[0,1]$ in the case of $\blue{M_p(1 - A_{\mu_\rho})}$. The values for $\E$ and $\ess \sup$ follow directly from the fact that $D_p$ satisfies Axiom~\ref{ax:normalization} by Theorem~\ref{thm:r_formula_mu} and the observation that $\cvar_0 = \E$ and $\cvar_1 = \ess \sup$.
    \qed
\end{proof}
    
Proposition~\ref{prop:bounds} shows that $D_p$ maps every spectral risk measure $\rho \in \mathcal{R}$ to a number on the interval between zero and one. This interval can be interpreted as the interval between risk-neutrality ($D_p(\E) = 0$) and maximal risk-aversion ($D_p(\ess \sup) = 1$).

\color{black} %\color{blue}
Next, we show that $D_p(\rho)$ is non-increasing in $p$ and bounded by the essential infimum and supremum of $\mu_\rho$.

\begin{proposition}\label{prop:D_p_limits}
    Let $\rho \in \mathcal{R}$ be given. Then, the following statements hold.
    \begin{enumerate}[(i)]
        \item The mapping $p \mapsto D_p(\rho)$ is non-increasing. In particular, for any $p,q \in \R$ with $p < q$, we have
        \begin{align}
            D_p(\rho) \geq D_q(\rho). \label{eq:D_p_monotonicity}
        \end{align}
        Moreover, if $0 < p < q$, then $D_p(\rho) = D_q(\rho)$ if and only if $\rho = \cvar_\alpha$ for some $\alpha \in [0,1]$.
        \label{prop:D_p_limits:monotonicity}
         \item Let $\underline{\alpha} := \operatorname*{ess\,inf}_{\mu_\rho}\alpha$ and $\overline{\alpha} := \operatorname*{ess\,sup}_{\mu_\rho}\alpha$. Then, 
         \begin{align}
             \lim_{p\to\infty}D_p(\rho) = \underline{\alpha}, \quad \text{and} \quad \lim_{p\to-\infty}D_p(\rho) = \overline{\alpha}. \label{eq:limits}
         \end{align} 
         Consequently, for every $p\in\mathbb{R}$,
        \begin{align}
            \underline{\alpha}
            \leq
            D_p(\rho)
            \leq
            \overline{\alpha}. \label{eq:boundaries}
        \end{align}
        \label{prop:D_p_limits:limits}
    \end{enumerate}
\end{proposition}
\begin{proof}
    See Appendix~\ref{sec:proofs}.
\end{proof}

\subsection{Identification from the degree profile}

Next, we show that a spectral risk measure is uniquely defined by its $p$-degrees, $p \in \naturalnumbers$. That is, two risk measures with the same $p$-degrees must be equal.

\begin{theorem} \label{thm:hausdorff}
    A spectral risk measure $\rho \in \mathcal{R}$ is uniquely defined by its $p$-degrees, $p \in \mathbb{N}$. That is, if $\rho_1, \rho_2 \in \mathcal{R}$ are such that $D_p(\rho_1) = D_p(\rho_2)$, for all $p \in \mathbb{N}$,  then, $\rho_1 = \rho_2$.
\end{theorem}
\begin{proof}
    Write $\mu_i$ for the measure from $\rho_i$'s Kusuoka representation, $i=1,2$. For every $p \in \mathbb{N}$, we have
    \begin{align*}
        D_p(\rho_1) = D_p(\rho_2) &\iff (1-D_p(\rho_1))^p = (1-D_p(\rho_2))^p \\
        &\iff \blue{M_p(1-A_{\mu_1})} = \blue{M_p(1-A_{\mu_2})} \\
        &\iff \int_{[0,1)} (1-\alpha)^p d\mu_1(\alpha) = \int_{[0,1)} (1-\alpha)^p d\mu_2(\alpha) \\
        &\iff \int_{[0,1]} \beta^p d\bar{\mu}_1(\beta) = \int_{[0,1]} \beta^p d\bar{\mu}_2(\beta),
    \end{align*}
    where, for $i=1,2$, $\bar{\mu}_i$ is the probability measure on the interval $[0,1]$ defined through $\bar{\mu}_i(E) = \mu_i(\bar{E})$, for all $E \in \mathcal{B}_{[0,1]}$, where $\bar{E} = \{1 - x \ : \ x \in E\}$. Thus, all $p$-moments of $\bar{\mu}_1$ and $\bar{\mu}_2$ are equal. By the Hausdorff moment problem  \cite{schmudgen2017moment}, it follows that $\bar{\mu}_1 = \bar{\mu}_2$, which implies that $\mu_1 = \mu_2$. As $\rho_i$ is uniquely defined through $\mu_i$, $i=1,2$, it follows that $\rho_1 = \rho_2$. \qed
\end{proof}

Theorem~\ref{thm:hausdorff} shows that the degrees $D_p(\rho)$, $p \in \naturalnumbers$, provide enough information to uniquely identify the spectral risk measure $\rho \in \mathcal{R}$. Combined with Theorem~\ref{thm:equivalence}, this yields the following corollary.

\begin{corollary}
    For any $\theta > 0$, a spectral risk measure $\rho \in \mathcal{R}$ is uniquely defined by the values $\bar{\rho}(Z_{p,\theta})$, $p \in \naturalnumbers$, where $\bar{\rho}$ is the truncation of $\rho$ from \eqref{eq:rho_truncated} and $Z_{p,\theta}$ is a random variable with a generalized Pareto distribution with shape parameter $-p$, scale parameter $1/\theta$, and location parameter $0$, $p \in \naturalnumbers$.
\end{corollary}
\begin{proof}
    The result follows immediately from Theorem~\ref{thm:hausdorff} and the representation of $D_p(\rho)$ in terms of $\bar{\rho}(Z_{p,\theta})$ for all $p > -1$.
\end{proof}

\color{black}

\color{black} %\color{blue}
\section{\blue{Consistency of $D_p$ with dominance relations}} \label{sec:dominance}
In this section, we analyze whether the degree functional $D_p$ is consistent with different concepts of dominance between the dual utility functions associated with spectral risk measures. That is, we explore whether $w_1$ dominating $w_2$ implies that $D_p(\rho_1) \geq D_p(\rho_2)$. We use two different notions of dominance: \textit{stochastic dominance} and \textit{dominance in relative-convexity order}. 

\subsection{Dominance relations}

We first define stochastic dominance, which is a way to compare random variables in terms of their distributions. In this paper, we are interested in comparing the distributions induced by dual utility functions $w_\rho$, $\rho \in \mathcal{R}$,  when interpreted as a cdf on the interval $[0,1]$. We emphasize that we are \textit{not} interested in stochastic dominance between random loss variables $Z \in \mathcal{Z}$ here, but merely between distributions induced by dual utility functions $w_\rho$. To simplify notation, we define stochastic dominance as a relation between the cdfs themselves, rather than the usual definition in terms of the random variables that have these cdfs. Moreover, we allow for stochastic dominance of fractional orders $r \in [1,\infty)$, in the spirit of Fishburn~\cite{fishburn1976continua}. 

\begin{definition}\label{def:stochastic_dominance}
    Let $\rho_1,\rho_2 \in \mathcal{R}$ be two spectral risk measures with dual utility functions $w_1$ and $w_2$, respectively. Then, for $k \in [1,\infty)$, we say that $w_1$ stochastically dominates $w_2$ in the $k$th order if 
    \begin{align}
        \int (t - x)_+^{k-1} dw_1(x) \leq \int (t - x)_+^{k-1} dw_2(x), \quad \forall t \in \R, \label{eq:stoch_dom}
    \end{align}
    where for $k=1$ we use the convention $(t - x)_+^0 := \mathbbm{1}_{x < t}$. %Moreover, for $r=0$, we say that $w_1$ stochastically dominates $w_2$ in the zeroth order if $X \geq Y$ a.s. For all $r \in \mathbb{Z}_+$, 
    We denote this relation by $w_1 \succeq^{(k)} w_2$. 
\end{definition}

It should be noted that stochastic dominance in $\bar{k}$th order implies stochastic dominance in $k$th order for all $k \geq \bar{k}$ \cite{fishburn1976continua}, so lower-order stochastic dominance relations are stronger. An important special case is first-order stochastic dominance, for which we obtain
\begin{align}
    w_1 \succeq^{(1)} w_2 \qquad \iff \qquad w_1 \leq w_2 \text{ uniformly.} \label{eq:fosd}
\end{align}
This means that $w_1$ assigns more probability mass to higher outcomes relative to $w_2$. We refer the reader to, e.g., Pflug and R\"{o}misch \cite{pflug2007modeling} for a more detailed discussion of stochastic dominance. 

Besides stochastic dominance, we use another concept of dominance between dual utility functions, called dominance in relative-convexity order. This notion was introduced by Palmer~\cite{palmer2003relative} and is based on the concept of \textit{relative convexity} between two functions by Jessen~\cite{jessen1931bemaerkninger}.

\begin{definition} \label{def:dominance}
    Let $\rho_1,\rho_2 \in \mathcal{R}$ be two spectral risk measures with dual utility functions $w_1$ and $w_2$, respectively. We say that $w_1$ dominates $w_2$ in \textit{relative-convexity order} if there exists a convex increasing function $h : [0,1] \to [0,1]$ such that $w_1 =  h \circ w_2$. We denote this relation by $w_1 \succeq^{(rc)} w_2$. 
\end{definition}

In the terminology of Palmer~\cite{palmer2003relative}, $w_1 \succeq^{(rc)} w_2$ means that $w_1$ is ``convex relative to'' $w_2$. Loosely speaking, we can interpret this by saying that $w_1$ is ``relatively more convex'' than $w_2$. As convexity of $w_\rho$ is equivalent to risk-aversion of $\rho$, this notion of relative convexity between dual utility functions seems particularly appropriate in the context of the degree functional $D_p$.

Now, before analyzing whether $D_p$ is consistent with the notions of dominance defined above, we first present a lemma that provides different characterizations of both notions of dominance and their relation.

\begin{lemma}\label{lemma:mother_dominance}
    Let $\rho_1,\rho_2 \in \mathcal{R}$ be two spectral risk measures with dual utility functions $w_1$ and $w_2$, respectively. Let $\text{id} : [0,1] \to [0,1]$ denote the identity function on $[0,1]$, defined by $\text{id}(x) = x$, $x \in [0,1]$. Finally, let $A_w$ be the (extended) Arrow-Pratt measure of absolute risk aversion from \eqref{eq:def_arrow_pratt}. Then, the following statements hold.
    \begin{enumerate}[(i)]
        \item $w_1 \succeq^{(1)} w_2$ if and only if $\rho_1(Z) \geq \rho_2(Z)$ for all $Z \in \mathcal{Z}$. \label{lemma:dom:unif_w}
        \item Suppose $w_1 = h \circ w_2$ for some $h: [0,1] \to [0,1]$. Then, 
        $w_1 \succeq^{(1)} w_2$ if and only if $h(x) \leq x$, $x \in [0,1]$. \label{lemma:dom:unif_circ}
        \item If $w_1 \succeq^{(rc)} w_2$, then $w_1 \succeq^{(1)} w_2$. \label{lemma:dom:implication}
        \item Assume that $w_1$ and $w_2$ are twice differentiable and strictly increasing. Then, $w_1 \succeq^{(rc)} w_2$ if and only if $A_{w_1} \geq A_{w_2}$ uniformly on $(0,1)$.\label{lemma:dom:arrow}
    \end{enumerate}
\end{lemma}
\begin{proof}
    See Appendix~\ref{sec:proofs}.
\end{proof}

Lemma~\ref{lemma:mother_dominance}\eqref{lemma:dom:unif_w} shows that first-order stochastic dominance between $w_1$ and $w_2$ is equivalent to uniform dominance between $\rho_1$ and $\rho_2$, which may be a natural criterion for $\rho_1$ being ``more risk averse'' than $\rho_2$. Thus, consistency with first-order stochastic dominance is a natural desirable property for $D_p$. We will explore this in Theorem~\ref{thm:consistent_dominance} below.

Lemma~\ref{lemma:mother_dominance}\eqref{lemma:dom:unif_circ} characterizes first-order stochastic dominance in terms of the function $h$ defining dominance in relative-convexity order. First-order stochastic dominance requires $h$ to be uniformly below the identity function $x \mapsto x$. Dominance in relative-convexity order, on the other hand, requires $h$ to be convex. As $h(0) = 0$ and $h(1) = 1$ due to the endpoint requirements on $w_1$ and $w_2$, the latter requirement is stronger, which is stated in Lemma~\ref{lemma:mother_dominance}\eqref{lemma:dom:implication}. 

Finally, Lemma~\ref{lemma:mother_dominance}\eqref{lemma:dom:arrow} relates dominance in relative-convexity order to the (extended) Arrow-Pratt measure of absolute risk aversion $A_w$. Recall that $A_w$ is a \textit{local} measure of the curvature of $w$. Lemma~\ref{lemma:mother_dominance}\eqref{lemma:dom:arrow} now states that $w_1 \succeq^{(rc)} w_2$ is equivalent to $w_1$ having a stronger local curvature than $w_2$ uniformly. It is natural to say that this implies that $\rho_1$ is more risk averse than $\rho_2$, and that it would be desirable if this were reflected as $D_p(\rho_1) \geq D_p(\rho_2)$. In other words, it would be desirable if the degree functional $D_p$ were consistent with dominance in relative-convexity order.

\color{black}

\color{black} %\color{blue}
\subsection{Consistency results}

\blue{We now show that the degree functional $D_p$ is consistent with dominance in relative-convexity order between dual utility functions for all $p \in \R$ and consistent with $k$th-order stochastic dominance between dual utility functions if $p \geq k-1$.}
%We can relate dominance to the degree of the risk measure. 

\color{black} %\color{blue}
\begin{theorem}\label{thm:consistent_dominance}
    Let $\rho_1, \rho_2 \in \mathcal{R}$ be two spectral risk measures with dual utility functions $w_1$ and $w_2$, respectively. Then, the following three statements hold.
    \begin{enumerate}[(i)]
        \item  If $w_1 \succeq^{(rc)} w_2$, then 
        \begin{align}
            D_p(\rho_1)\geq D_p(\rho_2), \quad p \in \mathbb{R}. \label{eq:consitency_rc}
        \end{align}
        \label{thm:consistent_dom:rc}
        \item If $w_1 \succeq^{(1)} w_2$, then 
        \begin{align}
            D_p(\rho_1)\geq D_p(\rho_2), \qquad p \geq - 1.   \label{eq:consitency_fosd}
        \end{align}
        \label{thm:consistent_dom:fosd}
        \item If $w_1 \succeq^{(k)} w_2$ for $k > 1$, then 
        \begin{align}
            D_p(\rho_1)\geq D_p(\rho_2), \qquad p \geq k - 1.   \label{eq:consitency_kosd}
        \end{align}
        \label{thm:consistent_dom:kosd}
    \end{enumerate}
\end{theorem}
\begin{proof}
    See Appendix~\ref{sec:proofs}.
\end{proof}

\color{black}

\color{black} %\color{blue}
Theorem~\ref{thm:consistent_dominance}\eqref{thm:consistent_dom:rc} shows that the degree functional $D_p$ is consistent with dominance in relative-convexity order: if one dual utility function is more convex than another in this relative sense, then its associated risk measure has a weakly higher $p$-degree for every $p \in \R$. This suggests that we can indeed interpret $D_p$ as a measure of the ``degree of convexity'' of the dual utility function. 

Consistency of $D_p$ with stochastic dominance depends on the order $k$.  Theorem~\ref{thm:consistent_dominance}\eqref{thm:consistent_dom:kosd} shows that higher values of $p$ ensure consistency with stochastic dominance of a wider range of orders $k > 1$. At $k=1$, however,  Theorem~\ref{thm:consistent_dominance}\eqref{thm:consistent_dom:fosd} establishes consistency whenever $p \geq -1$, which is a wider range than the bound $p \geq k-1$ for $k > 1$ would suggest. This structural break reflects the qualitatively different nature of first-order stochastic dominance: rather than ordering expectations only for a class of functions satisfying particular higher-order curvature restrictions, it orders expectations of all monotone transformations. Moreover, first-order stochastic dominance between the dual utility functions is equivalent to uniform dominance between the associated risk measures. As argued above, respecting this ordering is a natural requirement for $D_p$ to provide a reasonable measure of the degree of risk aversion. From this perspective, $p=-1$ constitutes a natural lower bound when consistency with first-order stochastic dominance is regarded as a desirable requirement. The example below shows that the lower bound is sharp: for every $p < -1$, $D_p$ may fail to be consistent with first-order stochastic dominance.

%Proposition~\ref{prop:consistent_with_dominance} and Corollary~\ref{cor:dominance_convexity} can be interpreted as logical validation of the degree functional $D_p$ for $p \geq -1$. If, for $\rho_1,\rho_2 \in \mathcal{R}$, $\rho_1$ uniformly dominates $\rho_2$, then it is natural to say that $\rho_1$ is ``more risk averse'' than $\rho_2$. Proposition~\ref{prop:consistent_with_dominance} shows that for $p \geq -1$, this is indeed reflected in the relative values $D_p(\rho_1)$ and $D_p(\rho_2)$. Interestingly, however, the result generally breaks down for $p < -1$, as is shown in the following counterexample. 

\begin{example}
    Let $p < -1$ be given and define Kusuoka measures $\mu_1=\frac{3}{10}\delta_{3/5}+\frac{7}{10}\delta_{4/5}$ and $\mu_2=\frac{1}{5}\delta_{1/5}+\frac{4}{5}\delta_{4/5}$. The corresponding dual utility functions are $w_1(t)=\frac{3}{4}\left(t-\frac{3}{5}\right)_+ +\frac{7}{2}\left(t-\frac{4}{5}\right)_+$, $w_2(t)=\frac{1}{4}\left(t-\frac{1}{5}\right)_+ +4\left(t-\frac{4}{5}\right)_+$. Direct verification gives $w_1(t)\leq w_2(t)$ for every $t\in[0,1]$; hence, $\rho_1$ uniformly dominates $\rho_2$. Next, write $I_p(\mu_i) = \int_{[0,1]} (1-\alpha)^p d\mu_i(\alpha)$. We have $I_p(\mu_1) = \frac{3}{10} \left(\frac{2}{5}\right)^p + \frac{7}{10}\left(\frac{1}{5}\right)^p$ and $I_p(\mu_2) = \frac{1}{5} \left(\frac{4}{5}\right)^p + \frac{4}{5}\left(\frac{1}{5}\right)^p$. Thus, $I_p(\mu_2) - I_p(\mu_1) = \frac{1}{5} \left(\frac{4}{5}\right)^p + \frac{1}{10}\left(\frac{1}{5}\right)^p - \frac{3}{10} \left(\frac{2}{5}\right)^p = \frac{1}{5 \cdot 5^p} ( 4^p + \frac{1}{2} 1^p - \frac{3}{2} 2^p) = 5^{-(p+1)} (x^2 - \frac{3}{2} x + \frac{1}{2}) = 5^{-(p+1)} (x-1)(x-\frac{1}{2})$, where $x = 2^p$. Since $p < -1$, we have $x = 2^p < 1/2$. By Theorem~\ref{thm:r_formula_mu}, $D_p(\rho_i) = 1 - (I_p(\mu_i))^{1/p}$, $i=1,2$. Since $y \mapsto 1 - y^{1/p}$ is increasing for $p < 0$, it follows that $D_p(\rho_1) < D_p(\rho_2)$. Thus, $D_p$ is not consistent with uniform dominance for $p < -1$.
\end{example}

\color{black}

\section{Extension to law invariant coherent risk measures} \label{sec:coherent}

\blue{We end by asking the question how the degree functional $D_p$ defined for spectral risk measures $\rho \in \mathcal{R}$ could be extended to the more general space $\bar{\mathcal{R}}$ of law-invariant coherent risk measures. To answer this question, we use the (minimal) Kusuoka representation from \eqref{eq:kusuoka_coherent}. As $\rho \in \bar{\mathcal{R}}$ is the supremum over the spectral risk measures $\rho_\mu$ associated with $\mu \in \mathcal{M}_\rho$, a reasonable requirement is that the degree of $\rho$ should be at least as high as the degree of each $\rho_\mu$. It is therefore natural to define $D_p(\rho)$ as the \textit{lowest} number that satisfies this condition for all $\mu \in \mathcal{M}_\rho$, i.e., the supremum of $D_p(\rho_{\mu})$ over all $\mu \in \mathcal{M}_\rho$.}

\begin{definition}\label{def:D_p_coherent}
    Consider a law invariant coherent risk measure $\rho \in \bar{\mathcal{R}}$. \blue{Let $\mathcal{M}_\rho$ be the set of probability measures in $\mathcal{P}([0,1])$ satisfying \eqref{eq:kusuoka_coherent} that is minimal in the sense of \cite{pichler2012uniqueness}.} 
    Then, for $p \in \R$, the $p$-degree (of risk aversion) of $\rho$ is defined by
    \begin{align*}
        D_p(\rho) = \sup_{\mu \in \mathcal{M}_\rho} \left\{ 1 - \blue{M_p(1 - A_\mu)} \right\}.
    \end{align*}
\end{definition}

\blue{This extension agrees with the original functional on the spectral subclass $\mathcal{R} \subseteq \bar{\mathcal{R}}$ and it preserves normalization for $\cvar_\alpha$, $\alpha \in [0,1]$, as the following proposition shows. }

\begin{proposition} \label{prop:generalization}
    Let $D_p : \bar{\mathcal{R}} \to \R$ be the extended degree functional from Definition~\ref{def:D_p_coherent}. Then, for any $\rho \in \mathcal{R} \subseteq \bar{\mathcal{R}}$, equations \eqref{eq:r_formula_w} and \eqref{eq:r_formula_mu} from Theorems~\ref{thm:r_formula_w} and \ref{thm:r_formula_mu} hold true. \blue{In particular, for $\rho = \cvar_\alpha \in \bar{\mathcal{R}}$, we obtain $D_p(\cvar_\alpha) = \alpha$, $\alpha \in [0,1]$, so $D_p$ satisfies the normalization Axiom~\ref{ax:normalization}.}
\end{proposition}
\begin{proof}
    \blue{As $\rho \in \mathcal{R}$, by \cite[Corollary~2.2]{pichler2012uniqueness} it has a Kusuoka representation in terms of a single measure $\mu_\rho \in \mathcal{P}([0,1])$, i.e., $\mathcal{M}_\rho = \{\mu_\rho\}$. Thus, by definition of $D_p$, we have $D_p(\rho) = \sup_{\mu \in \mathcal{M}_\rho} \left\{ 1 - \blue{M_p(1 - A_\mu)} \right\} = 1 - \blue{M_p(1 - A_{\mu_\rho})}$, which is the formula in \eqref{eq:r_formula_mu}. As $\rho \in \mathcal{R}$, $D_p(\rho)$ also equals the expression in \eqref{eq:r_formula_w} by Theorems~\ref{thm:r_formula_mu} and \ref{thm:r_formula_w}. For $\rho = \cvar_\alpha \in \bar{\mathcal{R}}$, it follows that $D_p(\cvar_\alpha) = 1 - M_p(1-A_{\mu_\rho}) = 1 - (1-\alpha) = \alpha$.
    }\qed
\end{proof}

\blue{The $p$-linearity in Axiom~\ref{ax:linearity} does not follow directly, however: although convex combinations of Kusuoka sets provide representations of convex combinations of risk measures, they need not coincide with the corresponding minimal Kusuoka sets used in Definition~\ref{def:D_p_coherent}. Likewise, an analogue of Axiom~\ref{ax:continuity} would require an appropriate notion of convergence for these set-valued representations, together with control of the relevant possibly unbounded integrals. Since identifying natural and sufficiently general formulations of these properties requires a separate analysis of minimal Kusuoka representations, we leave a systematic axiomatic study of the extended functional to future research.}

\section{Conclusion} \label{sec:conclusion}

\color{black} %\color{blue}
This paper introduces a family of functionals for quantifying the degree of risk aversion of spectral risk measures. The main value of this quantification is that it turns risk aversion from a qualitative property into a characteristic that can be compared across risk measures. The various representations developed in the paper show that this numerical meaning can be understood from several complementary perspectives, ranging from the curvature of the dual utility function to the tail behavior of the losses to which it is applied. The connection to generalized Pareto losses enables spectral risk measures to be calibrated according to their desired treatment of different tail behaviors.

The results also suggest several avenues for further research. A natural theoretical question is whether the extension to law-invariant coherent risk measures admits an axiomatic characterization comparable to that obtained for spectral risk measures. Finally, from an applied perspective, it would be interesting to further develop calibration procedures based on desired levels of conservatism toward different tail behaviors.

\color{black}
\begin{comment}
The degree functional $D_p$ formalizes what it means for a spectral or law invariant coherent risk measure to be ``more risk-averse'' than another. In this way, it extends the notion of global measures of risk aversion in the spirit of Arrow and Pratt from the primal world of expected utility functionals to the dual world of spectral risk measures, and to more general law invariant coherent risk measures. 

Thus, the degree functional $D_p$ offers a unifying language for comparing, ranking, and calibrating risk measures. This can be helpful in practice, when deciding which risk measure to use for a particular practical problem. Moreover, it opens the door for formalizing intuitive notions that some operations (e.g., mixing with another risk measure) make a risk measure ``more'' or ``less'' risk-averse. Indeed, future research might be aimed in this direction. Finally, future research could further explore the dual relationship between the degree functional $D_p$ for (dual) spectral risk measures and Arrow-Pratt-type measures of risk aversion for (primal) expected utility.
\end{comment}

\appendix 

\section{Proofs} \label{sec:proofs}

\color{black} %\color{blue}

In this section, we provide the proofs of Theorem~\ref{thm:r_formula_mu}, Theorem~\ref{thm:r_formula_w}, Theorem~\ref{thm:equivalence}, Proposition~\ref{prop:D_p_limits}, Lemma~\ref{lemma:mother_dominance}, and Theorem~\ref{thm:consistent_dominance}.

\begin{proof}[Proof of Theorem~\ref{thm:r_formula_mu}]
We prove existence and uniqueness of $D_p$ separately. To show existence, suppose that $D_p$ satisfies \eqref{eq:r_formula_mu}. We will prove that $D_p$ satisfies Axioms~\ref{ax:normalization}-\ref{ax:continuity}.

First, note that substituting \eqref{eq:r_formula_mu} in the definition of $s_p$ yields $s_p(\rho) = \int f_p d\mu_\rho$. Thus, for $\alpha \in [0,1]$, we have $s_p(\cvar_\alpha) = \int f_p(\beta) d\delta_\alpha = f_p(\alpha)$. Inverting the mapping $D_p(\rho) \mapsto s_p(\rho)$ yields $D_p(\cvar_\alpha) = f_p^{-1}(s_p(\cvar_\alpha)) = f_p^{-1}(f_p(\alpha)) = \alpha$. Thus, $D_p$ satisfies Axiom~\ref{ax:normalization}.

Next, let spectral risk measures $\rho_1,\ldots,\rho_m \in \mathcal{R}$ with associated Kusuoka measures $\mu_1,\ldots,\mu_m$ be given and let $\lambda_1,\ldots,\lambda_m$ be positive weights satisfying $\sum_{i=1}^m \lambda_i = 1$. Then, by linearity of the Kusuoka representation \eqref{eq:kusuoka_spectral} in $\mu_\rho$, it follows that for all $Z \in \mathcal{Z}$,
\begin{align*}
    \sum_{i=1}^m \lambda_i \rho_i(Z) = \sum_{i=1}^m \lambda_i \int_{[0,1]} \cvar_\alpha(Z) d\mu_i = \int_{[0,1]} \cvar_\alpha(Z) d\left( \sum_{i=1}^m \lambda_i \mu_i \right) = \rho_{\bar{\mu}}(Z),
\end{align*}
where $\rho_{\bar{\mu}}$ is the spectral risk measure with Kusuoka measure $\bar{\mu} := \sum_{i=1}^m \lambda_i \mu_i$. It follows that $s_p(\sum_{i=1}^m \lambda_i \rho_i) = s_p(\rho_{\bar{\mu}}) = \int f_p d\bar{\mu} = \int f_p d(\sum_{i=1}^m \lambda_i \mu_i) = \sum_{i=1}^m \lambda_i \int f_p d\mu_i = \sum_{i=1}^m \lambda_i s_p(\rho_i)$. Thus, $D_p$ satisfies Axiom~\ref{ax:linearity}.

Finally, suppose $\{\mu_n\}_{n=1}^\infty$ is a sequence of probability measures on $([0,1], \mathcal{B}_{[0,1]})$ that satisfy the conditions of Axiom~\ref{ax:continuity}. Then, by the second condition it follows immediately that $s_p(\rho_{\mu_n}) = \int f_p d\mu_n \to \int f_p d\mu = s_p(\rho_\mu)$, and thus that $D_p(\rho_{\mu_n}) \to D_p(\rho_\mu)$.\footnote{\blue{Note that weak convergence $\mu_n \Rightarrow \mu$ alone is not sufficient to guarantee that $s_p(\rho_{\mu_n}) \to s_p(\rho_\mu)$, since $f_p$ is then not bounded at $\alpha = 1$. This is the reason why we restrict to sequences satisfying $\int f_p d\mu_n \to \int f_p d\mu$ in Axiom~\ref{ax:continuity}.}} Thus, $D_p$ satisfies Axiom~\ref{ax:continuity}.

Next, we prove uniqueness. We assume that $D_p$ satisfies Axioms~\ref{ax:normalization}-\ref{ax:continuity} and we show that it must be of the form \eqref{eq:r_formula_mu}. Equivalently, we will show that $s_p$ must be of the form $s_p(\rho) = \int f_p d\mu_\rho$. 

We start by observing that for $\alpha \in [0,1]$, Axiom~\ref{ax:normalization} implies that $s_p(\cvar_\alpha) = f_p(D_p(\cvar_\alpha)) = f_p(\alpha)$. Next, we generalize to any $\rho \in \mathcal{R}$ with associated Kusuoka measure $\mu$ by constructing an appropriate sequence $\{ \mu_n\}_{n=1}^\infty$ converging to $\mu$. To do so, define the function $T_n(\alpha) := \frac{\floor{2^n \alpha}}{2^n}$, $\alpha \in [0,1]$. Note that for every $\alpha \in [0,1]$, we have $T_n(\alpha) \uparrow \alpha$ as $n \to \infty$. Define $\mu_n$ as the push-forward measure $\mu_n := \mu \circ T_n^{-1}$. Since each $\mu_n$ is finitely supported, we can write $\mu_n = \sum_{i=1}^{m_n} \lambda_{n,i} \delta_{\alpha_{n,i}}$ for appropriate values $\alpha_{n,i} \in [0,1]$, $i=1,\ldots,m_n$, and positive weights $\lambda_{n,i}$, $i=1,\ldots,m_n$, satisfying $\sum_{i=1}^{m_n} \lambda_{n,i} = 1$. By Axiom~\ref{ax:linearity}, it follows that
\begin{align}
    s_p(\rho_{\mu_n}) = s_p(\sum_{i=1}^{m_n} \lambda_{n,i} \cvar_{\alpha_{n,i}}) = \sum_{i=1}^{m_n} \lambda_{n,i} s_p(\cvar_{\alpha_{n,i}}) = \sum_{i=1}^{m_n} \lambda_{n,i} f_p(\alpha_{n,i}) = \int f_p d\mu_n. \label{eq:s_finite_mixing}
\end{align}

Next, we will use Axiom~\ref{ax:continuity} to pass to the limit as $n \to \infty$ in \eqref{eq:s_finite_mixing} and prove that $s_p(\rho_{\mu}) = \int f_p d\mu$. First, we show that the sequence $\{\mu_n\}_{n=1}^\infty$ satisfies the two conditions of Axiom~\ref{ax:continuity}. To prove weak convergence $\mu_n \Rightarrow \mu$, let $g : [0,1] \to \R$ be any continuous bounded test function. Then, as $g$ is continuous, $T_n(\alpha) \to \alpha$ implies that $g(T_n(\alpha)) \to g(\alpha)$, $\alpha \in [0,1]$. Since $g$ is bounded, the dominated convergence theorem implies that $\int g d\mu_n = \int g(T_n(\alpha)) d\mu(\alpha) \to \int g(\alpha) d\mu$, so $\mu_n \Rightarrow \mu$. To prove the second condition of Axiom~\ref{ax:continuity}, i.e., that
\begin{align}
    \int f_p d\mu_n \to \int f_p d\mu \label{eq:condition_f}
\end{align}
(i.e., that $\int f_p d\mu_n \to \int f_p d\mu$), we need to distinguish between three cases for the value of $p \in \R$. First, if $p > 0$, then $f_p(\alpha) = (1-\alpha)^p$ is continuous and bounded, so \eqref{eq:condition_f} follows from $\mu_n \Rightarrow \mu$. Second, if $p < 0$, then $f_p(\alpha) = (1-\alpha)^p$ is non-negative, monotonically increasing, and infinite at $\alpha = 1$. Since $T_n(\alpha) \uparrow \alpha$, the monotone convergence theorem implies that $\int f_p d\mu_n = \int f_p(T_n(\alpha)) d\mu \uparrow \int f_p d\mu$. This remains valid if the last integral equals $+\infty$. Finally, if $p=0$, then $f_0(\alpha) = \log(1-\alpha)$, $\alpha \in [0,1]$. Write $h(\alpha) := - f_0(\alpha)$, $\alpha \in [0,1]$. Then, $h$ is non-negative, monotonically increasing and infinite at $\alpha = 1$. By the monotone convergence theorem, $\int h d\mu_n = \int h(T_n(\alpha)) d\mu \uparrow \int h(\alpha) d\mu(\alpha)$. Multiplying by $-1$ on both sides yields $\int f_0 d\mu_n \downarrow \int f_0 d\mu$. This remains valid if the last integral equals $-\infty$. Combining the three cases, we have shown that for all $p \in \R$, $\mu_n \Rightarrow \mu$ and $\int f_p d\mu_n \to \int f_p d\mu$, so the sequence $\{\mu_n\}_{n=1}^\infty$ satisfies both conditions of Axiom~\ref{ax:continuity}. Thus, by Axiom~\ref{ax:continuity}, it follows that $D_p(\rho_{\mu_n}) \to D_p(\rho_\mu)$, which implies that 
\begin{align}
    s_p(\rho_{\mu_n}) \to s_p(\rho_\mu) \label{eq:limit_s}
\end{align}
Now, combining the limits in \eqref{eq:condition_f} and \eqref{eq:limit_s} with \eqref{eq:s_finite_mixing}, it follows that both limits must be equal, so that $s_p(\rho_\mu) = \int f_p d\mu$. This implies that $D_p(\rho_\mu) = f_p^{-1}(\int f_p d\mu) = 1 - M_{\mu}^p(1 - \alpha)$. Thus, the formula in \eqref{eq:r_formula_mu} is unique.
\qed
\end{proof}

\color{black}

\color{black} %\color{blue}

\begin{proof}[Proof of Theorem~\ref{thm:r_formula_w}]
By Lemma~\ref{lemma:identities}, the measure induced by $w_\rho$ is absolutely continuous on $[0,1)$, with density $w_\rho^{(r)}$, while an atom of $\mu_\rho$ at $1$ gives rise to an atom of the same mass in $dw_\rho$ at $1$. In particular, for $t \in [0,1)$, we have $w_\rho^{(r)}(t) = \int_{[0,t]} \frac{1}{1-\alpha} \, d\mu_\rho(\alpha)$.

First, consider $p>-1$ with $p\neq 0$. Using the expression for $w^{(r)}$ above, we obtain
\begin{align*}
    \int_{[0,1]} (1-t)^p \, dw_\rho(t)
    &=
    \int_{[0,1)} (1-t)^p
    \left(
        \int_{[0,t]} \frac{1}{1-\alpha} \, d\mu_\rho(\alpha)
    \right) dt
    +(1-1)^p\mu_\rho(\{1\}).
\end{align*}
Here and below, the expressions at the endpoint are understood in the extended-real sense. By Tonelli's theorem, we may interchange the order of integration, which yields
\begin{align*}
    \int_{[0,1]} (1-t)^p \, dw_\rho(t) &= \int_{[0,1)} \frac{1}{1-\alpha} \left( \int_\alpha^1 (1-t)^p \, dt \right) d\mu_\rho(\alpha) +(1-1)^p\mu_\rho(\{1\}).
\end{align*}
For every $\alpha<1$, the inner integral is $\int_\alpha^1 (1-t)^p \, dt = \frac{(1-\alpha)^{p+1}}{p+1}$, and therefore $\frac{1}{1-\alpha} \int_\alpha^1 (1-t)^p \, dt = \frac{(1-\alpha)^p}{p+1}$. Consequently, $(p+1)\int_{[0,1]} (1-t)^p \, dw_\rho(t) =
\int_{[0,1]} (1-\alpha)^p \, d\mu_\rho(\alpha)$. Notice that this also accounts for a possible atom of $\mu_\rho$ at $1$: its contribution is zero when $p>0$ and infinite when $-1<p<0$. Substituting the preceding identity into Theorem~1 yields $D_p(\rho) = 1 - \left( (p+1)\int_{[0,1]} (1-t)^p \, dw_\rho(t) \right)^{1/p}$.

Next, consider $p=0$. Applying the same argument to the nonnegative function $-\log(1-t)$ and then multiplying by $-1$, we obtain
\begin{align*}
    \int_{[0,1]} \log(1-t) \, dw_\rho(t) &= \int_{[0,1)} \frac{1}{1-\alpha} \left( \int_\alpha^1 \log(1-t) \, dt \right) d\mu_\rho(\alpha) +\log(1-1)\mu_\rho(\{1\}).
\end{align*}
For $\alpha<1$, we have $\int_\alpha^1 \log(1-t) \, dt = (1-\alpha)\big(\log(1-\alpha)-1\big)$, so that $\frac{1}{1-\alpha} \int_\alpha^1 \log(1-t) \, dt = \log(1-\alpha)-1$. It follows that $\int_{[0,1]} \log(1-t) \, dw_\rho(t) =  \int_{[0,1]} \log(1-\alpha) \, d\mu_\rho(\alpha)-1$. Thus, Theorem~\ref{thm:r_formula_mu} gives $D_0(\rho) = 1 - \exp\left( 1+\int_{[0,1]} \log(1-t) \, dw_\rho(t) \right)$.

Next, consider $p=-1$. By Lemma~\ref{lemma:identities} and the convention that $1/(1-\alpha)$ is infinite at $\alpha=1$, we have $w_\rho^{(l)}(1) = \int_{[0,1]} \frac{1}{1-\alpha} \, d\mu_\rho(\alpha)$. In particular, if $\mu_\rho$ has positive mass at $1$, both sides are infinite. The $p=-1$ case of Theorem~\ref{thm:r_formula_mu} therefore gives $D_{-1}(\rho) = 1-\frac{1}{w_\rho^{(l)}(1)}$, where the reciprocal of infinity is understood to be zero.

Finally, consider $p<-1$. Recall that $\varphi_\rho(t) = w_\rho^{(l)}(t)$, and that by Lemma~\ref{lemma:identities}, for $t<1$, $\varphi_\rho(t) =\int_{[0,t)} \frac{1}{1-\alpha}\,d\mu_\rho(\alpha)$. Hence, on $[0,1)$, the Lebesgue--Stieltjes measure induced by $\varphi_\rho$ satisfies $d\varphi_\rho(t) = \frac{1}{1-t}\,d\mu_\rho(t)$. Therefore, for every $r<1$, we have $\int_{[0,r]}(1-t)^p\,d\mu_\rho(t) = \int_{[0,r]}(1-t)^{p+1}\,d\varphi_\rho(t)$. Letting $r\uparrow1$ and using monotone convergence gives $\int_{[0,1)}(1-t)^p\,d\mu_\rho(t) = \int_{[0,1)}(1-t)^{p+1}\,d\varphi_\rho(t)$. It remains to consider a possible atom of $\mu_\rho$ at $1$. If $\mu_\rho(\{1\})>0$, then the left derivative of $w_\rho$ at $1$ is infinite, that is, $\varphi_\rho(1) = +\infty$. Accordingly, the extended Lebesgue--Stieltjes measure induced by  $\varphi_\rho$ has infinite mass at the endpoint $1$. Since $p+1<0$, its contribution to the Stieltjes integral is infinite. At the same time, since $p<0$, an atom of $\mu_\rho$ at $1$ also makes the corresponding integral infinite. Thus, $\int_{[0,1]}(1-t)^p\,d\mu_\rho(t) = \int_{[0,1]}(1-t)^{p+1}\,d\varphi_\rho(t)$, where both integrals are understood in the extended-real sense.  Substituting this identity into Theorem~\ref{thm:r_formula_mu} yields $D_p(\rho) = 1- \left( \int_{[0,1]} (1-t)^{p+1}\,d\varphi_\rho(t) \right)^{1/p}$. This concludes the proof.
\qed
\end{proof}

\begin{proof}[Proof of Theorem~\ref{thm:equivalence}]
    First note that for all $m \in \naturalnumbers$, $Z_{p,\theta} \wedge m \in \mathcal{Z}$ and $\rho(Z_{p,\theta} \wedge m) = \int_{[0,1]} F_{Z_{p,\theta} \wedge m}^{-1}(t) dw_\rho(t) = \int_{[0,1]} ( F_{{p,\theta}}^{-1}(t) \wedge m) dw_\rho(t)$. Hence, by the monotone convergence theorem, 
    \begin{align}
        \bar{\rho}(Z_{p,\theta}) = \int_{[0,1]}  F_{{p,\theta}}^{-1}(t)  dw_\rho(t), \label{eq:truncation}
    \end{align}
    where $+\infty$ is allowed. 
    
    For $p \neq 0$, the quantile function of $Z_{p,\theta}$ is given by the inverse of $F_{p,\theta}$, i.e., $F_{Z_{p,\theta}}^{-1}(t) = \frac{1 - (1 - t)^p}{\theta p}$, $t \in [0,1]$. Then, for any $\rho \in \mathcal{R}$, we have by \eqref{eq:truncation} that $\bar{\rho}(Z_{p,\theta}) = \int_0^1 F_{{p,\theta}}^{-1}(t) dw_\rho(t) = \int_0^1 \frac{\blue{1 - } (1 - t)^p}{\theta p} d w_\rho(t) = (\theta p)^{-1} \left(\blue{1 - } \int_0^1 (1 - t)^p dw_\rho(t)  \right)$. Therefore, $\int_0^1 (1-t)^p dw_\rho(t) = 1 - p \theta \bar{\rho}(Z_{p,\theta})$, and Theorem~\ref{thm:r_formula_w} gives the formula in \eqref{eq:D_formula_Z}.
    
    Next, suppose $p = 0$. Then, the quantile function of $Z_{p,\theta}$ is given by $F_{p,\theta}^{-1}(t) = \linebreak - \theta^{-1} \log(1 - t)$, $t \in [0,1]$, so $\bar{\rho}(Z_{p,\theta}) = \int_0^1 F^{-1}_{p,\theta}(t) dw_\rho(t) = - \theta^{-1} \int_0^1 \log(1 - t) d w_\rho(t)$, and Theorem~\ref{thm:r_formula_w} gives the formula in \eqref{eq:D_formula_Z}. \qed
\end{proof}

\begin{proof}[Proof of Proposition~\ref{prop:D_p_limits}]
    First, we prove \eqref{prop:D_p_limits:monotonicity}. Let $p < q$ and note that the monotonicity of generalized means \cite[Theorem~3]{witkowski2004new} implies that $M_p(1 - A_{\mu_\rho}+\varepsilon)\leq M_q(1 - A_{\mu_\rho}+\varepsilon)$ for every $\varepsilon>0$. (The additional $\varepsilon$ is necessary to account for a potential atom at $A_{\mu_\rho} = 1$, which invalidates the positivity assumption on the integrand in the referenced monotonicity theorem.) Letting $\varepsilon\downarrow0$ yields $M_p(1 - A_{\mu_\rho})\leq M_q(1 - A_{\mu_\rho})$, and hence $D_p(\rho) \geq D_q(\rho)$.  Here, the limit follows by dominated convergence when $p > 0$, by monotone convergence when $p < 0$, and for $p=0$ by the definition $M_0(1 - A_{\mu_\rho})=\exp(\mathbb{E}[\log (1 - A_{\mu_\rho})])$, allowing $\mathbb{E}[\log (1 - A_{\mu_\rho})]=-\infty$.

    Next, let $0<p<q$ and write $X_{\mu_\rho} := 1 - A_{\mu_\rho}$. Since $x\mapsto x^{q/p}$ is strictly convex, Jensen's inequality gives $\mathbb{E}[X_{\mu_\rho}^q] \geq \big(\mathbb{E}[X_{\mu_\rho}^p]\big)^{q/p}$, with equality if and only if $X_{\mu_\rho}^p$ is constant almost surely. Taking the power $1/q$ gives $M_q(X_{\mu_\rho})\geq M_p(X_{\mu_\rho})$, with equality if and only if $X_{\mu_\rho}$ is constant almost surely. Since $X_{\mu_\rho}=1-A_{\mu_\rho}$, this is equivalent to $\mu_\rho=\delta_\alpha$ for some $\alpha\in[0,1]$, which, by the Kusuoka representation, is equivalent to $\rho=\operatorname{CVaR}_\alpha$.

    For the endpoint limits, since $X_{\mu_\rho}\in L^\infty$, the standard limiting
    property of $L^p$-norms
    \cite[Chapter~1, Proposition~2.2]{stein2011functional} yields $\lim_{p\to\infty}M_p(X_{\mu_\rho}) =\operatorname*{ess\,sup}X_{\mu_\rho}  = 1-\underline{\alpha}$, and therefore $\lim_{p\to\infty}D_p(\rho) =\underline{\alpha}$. Next, write $b=\operatorname*{ess\,inf}X_{\mu_\rho}=1-\overline{\alpha}$. If $b>0$, then $X_{\mu_\rho}^{-1}\in L^\infty$ and $M_{-p}(X_{\mu_\rho})=1/M_p(X_{\mu_\rho}^{-1})$. Hence, $\lim_{p\to\infty}M_{-p}(X_{\mu_\rho}) =\frac{1}{\operatorname*{ess\,sup}X_{\mu_\rho}^{-1}} =b$. If $b=0$, then for every $\varepsilon>0$, $c_\varepsilon:=\mathbb{P}(X_{\mu_\rho}<\varepsilon)>0$. For $p>0$, $\mathbb{E}[X_{\mu_\rho}^{-p}]\geq c_\varepsilon\varepsilon^{-p}$ whenever the expectation is finite, and consequently $M_{-p}(X_{\mu_\rho})\leq c_\varepsilon^{-1/p}\varepsilon$. The same inequality holds if $\mathbb{E}[X_{\mu_\rho}^{-p}]=\infty$, since then $M_{-p}(X_{\mu_\rho})=0$. Thus, $\limsup_{p\to\infty}M_{-p}(X_{\mu_\rho}) \leq\varepsilon$. Since $\varepsilon>0$ is arbitrary, $\lim_{p\to\infty}M_{-p}(X_{\mu_\rho}) = 0 = b$. Therefore, in either case, $\lim_{p\to-\infty}M_p(X_{\mu_\rho}) =1-\overline{\alpha}$, so that $\lim_{p\to-\infty}D_p(\rho) =\overline{\alpha}$.

    Finally, monotonicity together with the two endpoint limits gives
    $\underline{\alpha}\leq D_p(\rho)\leq\overline{\alpha}$ for every
    $p\in\mathbb{R}$.
    \qed
\end{proof}

\begin{proof}[Proof of Lemma~\ref{lemma:mother_dominance}]
    We first prove \eqref{lemma:dom:unif_w}. Suppose that $w_1 \succeq^{(1)} w_2$, so that $w_1 \leq w_2$ uniformly by \eqref{eq:fosd}. Let $Z \in \mathcal{Z}$ be given. Then, since $F_Z^{-1}$ is non-decreasing, we obtain
    \begin{align*}
        \rho_1(Z) &= \int_0^1 F_{Z}^{-1}(u) dw_1(u) \geq \int_0^1 F_{Z}^{-1}(u) dw_2(u) = \rho_2(Z),
    \end{align*}
    Next, suppose that $\rho_1(Z) \geq \rho_2(Z)$ for all $Z \in \mathcal{Z}$. For $\bar{u} \in [0,1]$, let $Z_{\bar{u}}$ be a Bernoulli random variable with success probability $1 - \bar{u}$. Then, $F_{Z_{\bar{u}}}^{-1}(u) = \mathbbm{1}_{(u > \bar{u})}(u)$. By dominance of $\rho_1$ over $\rho_2$ we get
    \begin{align*}
        \rho_1(Z_{\bar{u}}) \geq \rho_2(Z_{\bar{u}}) &\iff \int_0^1 \mathbbm{1}_{(u > \bar{u})}(u) dw_1(u) \geq \int_0^1 \mathbbm{1}_{(u > \bar{u})}(u) dw_2(u) \\
        & \iff w_1(1) - w_1(\bar{u}) \geq w_2(1) - w_2(\bar{u}) \\
        & \iff w_1(\bar{u}) \leq w_2(\bar{u}),        
    \end{align*}
    where we used that $w_1(1) = w_2(1) = 1$. Since $\bar{u} \in [0,1]$ is chosen arbitrarily, we obtain $w_1(u) \leq w_2(u)$ for all $u \in [0,1]$, so $w_1 \leq w_2$ uniformly, which is equivalent to $w_1 \succeq^{(1)} w_2$ by \eqref{eq:fosd}.

    To prove \eqref{lemma:dom:unif_circ}, note that $w_1 \succeq^{(1)} w_2 \iff  w_1(t) \leq w_2(t), \ t \in [0,1] \iff h(w_2(t)) \leq w_2(t), \ t \in [0,1]$. Since $w_2$ is continuous with $w_2(0) = 0$ and $w_2(1) = 1$, its range is all of $[0,1]$. Hence, the above is equivalent to $h(x) \leq x$, $x \in [0,1]$. 

    To prove \eqref{lemma:dom:implication}, note that $h$ is convex and must satisfy $h(0)=0$ and $h(1)=1$ due to the endpoint values of $w_1$ and $w_2$. Thus, $h(x) \leq x$, and the result follows by \eqref{lemma:dom:unif_circ}.

    Finally, to prove \eqref{lemma:dom:arrow}, recall that $w_1 = h \circ w_2$. Differentiating yields $w_1'(t) = h'(w_2(t))w_2'(t)$, so $h'(w_2(t)) = \frac{w_1'(t)}{w_2'(t)}$. Since $h$ is convex, $h'$ is non-decreasing. Since $w_2$ is increasing, $h'(w_2(t))$ therefore non-decreasing. Differentiating gives $\frac{d}{dt} h'(w_2(t)) = \frac{d}{dt} \frac{w_1'(t)}{w_2'(t)} = \frac{w_2'(t) w_1''(t) - w_1'(t) w_2''(t)}{(w_2'(t))^2} = \frac{w_1'(t)}{w_2'(t)} \left(\frac{w_1''(t)}{w_1'(t)} - \frac{w_2''(t)}{w_2'(t)}\right) \geq 0$. Since $w_1'/w_2' > 0$ on $(0,1)$, this implies $\frac{w_1''(t)}{w_1'(t)} - \frac{w_2''(t)}{w_2'(t)} \geq 0$, i.e., $A_{w_1}(t) \geq A_{w_2}(t)$, $t \in (0,1)$. Conversely, suppose that $A_{w_1}(t) \geq A_{w_2}(t)$, $t \in (0,1)$. Define $h = w_1 \circ w_2^{-1}$. Then, $w_1 = h \circ w_2$, and $h'(w_2(t)) = \frac{w_1'(t)}{w_2'(t)}$. Moreover, for $t \in (0,1)$, we have using the derivation above that $\frac{d}{dt} \frac{w_1'(t)}{w_2'(t)} =  \frac{w_1'(t)}{w_2'(t)} (A_{w_1}(t) - A_{w_2}(t)) \geq 0$. Thus, $h'(w_2(t))$ is non-decreasing in $t$. Since $w_2$ is strictly increasing, $h'$ is non-decreasing, so $h$ is convex. Hence, $w_1 = h \circ w_2$ for a convex $h$, so $w_1 \succeq^{(rc)} w_2$.    
    %again consider $h = w_1 \circ w_2^{-1}$. For any $x \in (0,1)$, we have by the chain rule that $h'(x) = w_1'(w_2^{-1}(x))(w_2^{-1})'(x)$. By the inverse-function rule, $(w_2^{-1})'(x) = \frac{1}{w_2'(w_2^{-1}(x))}$. Substituting this yields $h'(x) = \frac{w_1'(w_2^{-1}(x))}{w_2'(w_2^{-1}(x))}$. This yields $h'(w_2(t)) = \frac{w_1'(t)}{w_2'(t)}$ for any $t \in [0,1]$. Convexity of $h$ implies that $h'$ is non-decreasing. As $w_2$ is non-decreasing, this implies that the mapping $t \mapsto h'(w_2(t)) = \frac{w_1'(t)}{w_2'(t)}$ must be non-decreasing. That is, its derivative must be non-negative, i.e., $\frac{d}{dt} \frac{w_1'(t)}{w_2'(t)} = \frac{w_1''(t) w_2'(t) - w_1'(t) w_2''(t)}{(w_2'(t))^2} \geq 0$. As $w_2'(t) \geq 0$, this implies that $\frac{w_1''(t)}{w_1'(t)} \geq \frac{w_2''(t)}{w_2'(t)}$, i.e., $A_{w_1}(t) \geq A_{w_2}(t)$ for all $t \in [0,1]$. 
    \qed
\end{proof}

\begin{proof}[Proof of Theorem~\ref{thm:consistent_dominance}]
    First, we prove \eqref{thm:consistent_dom:rc}. Let $\mu_i$ be the Kusuoka measure associated with $\rho_i$, $i=1,2$. Using Lemma~\ref{lemma:identities}, we obtain $w_i(t)+(1-t)w_i^{(r)}(t) = \int_{[0,t]} \frac{t-\alpha+1-t}{1-\alpha}\,d\mu_i(\alpha) = \mu_i([0,t]) =: F_{\mu_i}(t)$, $t \in [0,1)$.  At $t=1$, the same identity holds in the form $F_{\mu_i}(1)=w_i(1)=1$.

    We now show that $F_{\mu_1}\leq F_{\mu_2}$ uniformly. Fix $t<1$ and set $x:=w_2(t)$ and $c:=(1-t)w_2^{(r)}(t)$. Since $w_1=h\circ w_2$, the chain rule for right derivatives yields $w_1^{(r)}(t)=h^{(r)}(x)w_2^{(r)}(t)$. Consequently, $F_{\mu_1}(t) = h(x)+c h^{(r)}(x)$ and $F_{\mu_2}(t) = x+c$. Convexity of $w_2$ and $w_2(1)=1$ imply $1\geq w_2(t)+(1-t)w_2^{(r)}(t)=x+c$, and therefore $c\leq1-x$. Moreover, convexity of $h$, together with $h(0)=0$ and $h(1)=1$, implies $h(x)\leq x$.  If $h^{(r)}(x)\leq1$, these two inequalities immediately give $h(x)+c h^{(r)}(x)\leq x+c$. If $h^{(r)}(x)>1$, convexity of $h$ gives $1-h(x) \geq (1-x)h^{(r)}(x)$, and hence $x-h(x)\geq(1-x)(h^{(r)}(x)-1) \geq c(h^{(r)}(x)-1)$. This is again equivalent to $h(x)+c h^{(r)}(x)\leq x+c$. Thus, $F_{\mu_1}(t)\leq F_{\mu_2}(t)$ for every $t\in[0,1]$.

    It follows that $\mu_1$ first-order stochastically dominates $\mu_2$. Thus, if $A_i\sim\mu_i$, the variables may be coupled so that $A_1\geq A_2$ almost surely, and hence $1-A_1\leq1-A_2$ almost surely. Since the generalized mean is increasing in its argument for every $p\in\mathbb{R}$, we obtain $M_p(1-A_1)\leq M_p(1-A_2)$. Finally, using $D_p(\rho_i)=1-M_p(1-A_i)$ gives $D_p(\rho_1)\geq D_p(\rho_2)$. This proves \eqref{thm:consistent_dom:rc}.

    Next, we prove \eqref{thm:consistent_dom:fosd}. We will prove the result by contradiction. Suppose that $D_p(\rho_1) < D_p(\rho_2)$ for some $p \geq -1$. We will first consider the case $p > 0$. %Other cases will follow similarly. 
    We have by Theorem~\ref{thm:r_formula_w} that
    \begin{align*}
        &D_p(\rho_1) < D_p(\rho_2) \\
        \iff& 1 - \left[ (p + 1) \int_0^1 (1-t)^p dw_1(t) \right]^{1/p} < 1 - \left[ (p + 1) \int_0^1 (1-t)^p dw_2(t) \right]^{1/p} \\
        \iff& \left[ (p + 1) \int_0^1 (1-t)^p dw_1(t) \right]^{1/p} > \left[ (p + 1) \int_0^1 (1-t)^p dw_2(t) \right]^{1/p} \\
        \iff& \int_0^1 (1-t)^p dw_1(t) > \int_0^1 (1-t)^p dw_2(t).
    \end{align*}
    Recall from \eqref{eq:fosd} that $w_1 \succeq^{(1)} w_2$ implies that $w_1 \leq w_2$ uniformly. Also recall that both $w_1$ and $w_2$ are cdfs on $[0,1]$. So $w_1$ puts more probability mass on higher values for $t$ than $w_2$. Because $t \mapsto (1-t)^p$ is decreasing, it follows that $\int_0^1 (1-t)^p dw_1(t) \leq \int_0^1 (1-t)^p dw_2(t)$, which contradicts the inequality above. This proves the claim for the case $p > 0$. The proof for the case $-1 < p < 0$ is analogous, accounting for two additional minus signs that cancel out against each other. The proof for the case $p=0$ is also analogous, using the relevant log-exponential formula from Theorem~\ref{thm:r_formula_w}. To prove the case $p=-1$, note that by \eqref{eq:fosd}, we have $w_1 \leq w_2$ uniformly. Since $w_1(1) = w_2(1)$ and both $w_1$ and $w_2$ are convex, it follows that $\varphi_1(1) \geq \varphi_2(1)$. Thus, by Theorem~\ref{thm:r_formula_w}, we obtain $D_{-1}(\rho_1) = 1 - 1/\varphi_1(1) \geq 1 - 1/\varphi_2(1) = D_{-1}(\rho_2)$. 

    Finally, we prove \eqref{thm:consistent_dom:kosd}. By \cite[Theorem~1]{fishburn1976continua}, $w_1 \succeq^{(k)} w_2$ implies $w_1 \succeq^{(p)} w_2$ for all $p \geq k$. By Definition~\ref{def:stochastic_dominance}, we have $\int_0^1(t - x)_+^{p-1}dw_1(x) \leq \int_0^1 (t - x)_+^{p-1} dw_2(x)$,  $t \in \R$. Substituting $t=1$, we obtain 
    \begin{align}
        \int_0^1 (1 - x)^{p-1} dw_1(x) \leq \int_0^1 (1 - x)^{p-1} dw_2(x). \label{eq:comparing_one_minus}
    \end{align}
    For $p \geq k > 1$, we have by Theorem~\ref{thm:r_formula_w} that
    \begin{align}
        D_{p-1}(\rho_i) = 1 - \left[ p \int_0^1 (1 - t)^{p-1} dw_\rho(t) \right]^{\frac{1}{p-1}}. \label{eq:D_p_minus_one}
    \end{align}
    Combining \eqref{eq:comparing_one_minus} and \eqref{eq:D_p_minus_one}, we obtain $\blue{D_{p-1}}(\rho_1) \geq \blue{D_{p-1}}(\rho_2)$.
    \qed
\end{proof}

\color{black} %\color{blue}

\color{black}

%% =======================
%% MAIN BODY ABOVE
%% =======================
\color{black} %\color{blue}
\begin{acknowledgements}
The author thanks two anonymous reviewers for their comments that substantially improved the contents of the paper. Furthermore, the author thanks Maikel Houbaer for insights into the relation between the degree functional $D_p$ in this paper and related concepts in the expected-utility literature. Finally, the author acknowledges the use of ChatGPT-5.6 for brainstorming, proof assistance, and copy editing.
%The authors thank Andrzej Ruszczy{\'n}ski, Darinka Dentcheva, and Tito Homem-de-Mello for valuable comments on earlier versions of this work.
%Possibly: Tito Homem-de-Mello if we use his idea for convergence for CVaR
\end{acknowledgements}
\color{black}

% Authors must disclose all relationships or interests that 
% could have direct or potential influence or impart bias on 
% the work: 
%
%\section*{Conflict of interest}
%The authors declare that they have no conflict of interest.

% BibTeX users please use one of
%\bibliographystyle{spbasic}      % basic style, author-year citations
\bibliographystyle{spmpsci}      % mathematics and physical sciences
\bibliography{references}   % name your BibTeX data base

\end{document}